\documentclass{article}
\usepackage{alltt, amsmath,verbatim,moreverb,amssymb,color,bm,graphicx,sectsty,ulem}
\usepackage[margin=1.5in]{geometry}
\usepackage[english]{babel}
\usepackage[utf8]{inputenc}
\usepackage{fancyhdr}
\usepackage[T1]{fontenc}
\usepackage{titling}
\usepackage{bm}
\usepackage{setspace}
\usepackage{comment}
\usepackage{caption}
\usepackage{xcolor}
\usepackage{ulem} 

\usepackage{algorithmicx}
\usepackage[noend]{algpseudocode}
\usepackage{algorithm}

\usepackage{amsthm}
\usepackage{amsmath}
\usepackage{natbib}
\usepackage[colorlinks,citecolor=blue,urlcolor=blue,filecolor=blue,backref=page]{hyperref}
\usepackage{cleveref}
\usepackage{graphicx}

\renewcommand{\hat}[1]{\widehat{#1}}

\usepackage{authblk}
\title{Bayesian Projection Pursuit Regression}
\author[1,2]{Gavin Collins}
\author[1]{Devin Francom}
\author[1]{Kellin Rumsey}
\affil[1]{Los Alamos National Laboratory}
\affil[2]{The Ohio State University}
\date{\vspace{-5ex}}                     

\usepackage{fancyhdr}
\begin{document}
\maketitle








\begin{abstract}
\noindent In projection pursuit regression (PPR), an unknown response function is approximated by the sum of $M$ ``ridge functions,'' which are flexible functions of one-dimensional projections of a multivariate input space. Traditionally, optimization routines are used to estimate the projection directions and ridge functions via a sequential algorithm, and $M$ is typically chosen via cross-validation. We introduce the first Bayesian version of PPR, which has the benefit of accurate uncertainty quantification. To learn the projection directions and ridge functions, we apply novel adaptations of methods used for the single ridge function case ($M=1$), called the Single Index Model, for which Bayesian implementations do exist; then use reversible jump MCMC to learn the number of ridge functions $M$. We evaluate the predictive ability of our model in 20 simulation scenarios and for 23 real datasets, in a bake-off against an array of state-of-the-art regression methods. Its effective performance indicates that Bayesian Projection Pursuit Regression is a valuable addition to the existing regression toolbox. 
\end{abstract}



\onehalfspacing
\section{Introduction}
The fundamental task of supervised regression learning is to approximate an unknown response function $f : \mathbb{R}^p \to \mathbb{R}$ given $n$ possibly noisy realizations $\bm y = (y_1,\dots,y_n)^\prime$ at input locations $\bm X = [\bm x_1 \ \dots \ \bm x_n]^\prime$, where $\bm x_i\in\mathbb{R}^p$ and $E[y_i] = f(\bm x_i)$ for $i = 1,\dots,n$. In likelihood-based methods, it is common to assume that $y_i \sim \mathcal{N}\left(f(\bm x_i), \sigma^2\right)$. A standard linear regression model will often suffice, but sometimes nonlinearities, inert dimensions of the input space, and/or complex interactions between input dimensions demand a more flexible framework. Existing tools range from likelihood-free methods like Neural Networks (NN; \cite{nn}), Random Forests (RF; \cite{rf}), Gradient Boosted Trees (GBT; \cite{gb}), Multivariate Adaptive Regression Splines (MARS; \cite{mars}), LASSO \citep{lasso}, and Projection Pursuit Regression (PPR; \newline\cite{ppr}); to probabilistic models like Bayesian Additive Regression Trees (BART; \cite{bart}), Bayesian MARS (BMARS; \cite{bmars}), Gaussian processes (GP; \cite{gp}) and approximate Gaussian processes, including sparse Gaussian processes \citep{sgp}, stochastic variational Gaussian processes \citep{vgp} and local approximate Gaussian processes (LAGP; \newline\cite{lagp}). Methods vary in their predictive accuracy, uncertainty quantification, interpretability, execution time, etc.

PPR has been part of the standard regression toolbox for over 40 years. In essence, PPR uses back-fitting to approximate $f$ with a sum of $M$ ``ridge functions,'' which are flexible functions of one-dimensional projections of the input space. Though less popular than, say, NN, RF, GBT, and LASSO, PPR is able to adapt to complex nonlinearities and interactions, with fast execution for small-to-moderate input dimension $p$. The strong presence of PPR as a standard regression tool is evidenced by its inclusion in the default-loaded R package \textit{stats} \citep{R}. 

Probabilistic models are equipped with uncertainty quantification, which is crucial in some applications. Furthermore, Bayesian probabilistic models implicitly average over many estimates of $f$ using draws from its posterior distribution to approximate it with maximal accuracy. In recent years, probabilistic versions of some standard likelihood-free models have been developed. For example, BART can be viewed as a Bayesian version of GBT, where $f$ is approximated by the sum of $M$ regression trees. BART uses a Bayesian back-fitting approach \citep{bbf}, which simplifies the MCMC algorithm so that each tree is fit to the residuals of the others, one at a time, in an iterative process. An update to an individual tree involves a Reversible Jump MCMC (RJMCMC; \cite{rjmcmc}) step to infer the tree structure, followed by a Gibbs step to infer the corresponding terminal node parameters. Another example is BMARS, which is a Bayesian version of Friedman's MARS model. In BMARS, RJMCMC is used to infer the number $M$ of basis functions, which are tensor products of one-dimensional first-order spline bases at learned knot points \citep{plate}. In both BART and BMARS, the RJMCMC steps are greatly simplified by integrating regression coefficients out of the likelihood. Fully Bayesian neural networks \citep{bayes_nn} have also been implemented, but computation tends to be prohibitively slow. Instead, variational Bayes has emerged as a standard way to approximate the marginal posterior distributions of NN parameters \citep{var_nn1,var_nn2}. 

To our surprise, we have not found an existing Bayesian version of PPR. This may be because it is challenging to learn the number of ridge functions $M$ in the model and to simultaneously estimate ridge functions and projection directions. There are, however, existing Bayesian versions of the Single Index Model (SIM; \cite{sim1,sim2,friedmanandtukey}), which is a PPR model with only a single ridge function (i.e. $M = 1$); but these are not flexible enough to model some complex functions $f$ with sufficient accuracy. Building on traditional PPR and current Bayesian versions of the SIM model, we propose an original, fully Bayesian Projection Pursuit Regression (BPPR) model, equipped with uncertainty quantification. We utilize RJMCMC techniques---as in BART and BMARS---to learn the number $M$ of ridge functions needed to model a particular response function $f$, and also offer novel innovations to facilitate greater predictive accuracy and faster execution. Our methodology is implemented in a publicly available R package (\url{https://github.com/gqcollins/BayesPPR}).

The remainder of the paper is outlined as follows. In \cref{sec:methods}, we describe the PPR and SIM models, introduce BPPR and its accompanying MCMC algorithm for real-valued input and response, and offer extensions to categorical input and multivariate response. In \cref{sec:app}, we demonstrate typical use of BPPR on simulated and real datasets, including interpretable inference and uncertainty quantification, and conduct a ``bake-off'' consisting of 20 simulated scenarios and 23 real datasets, where BPPR compares favorably to current state-of-the-art regression techniques in its predictive accuracy and uncertainty quantification. We summarize our work in \cref{sec:summary}.

\section{Methods}\label{sec:methods}

\subsection{Projection Pursuit Regression}

PPR models $f$ as
\begin{equation}\label{eq:ppr}
    f(\bm x) = \sum_{m=1}^M g_m(\bm x^\prime \bm\theta_m),
\end{equation}
\noindent where $g_1,\dots,g_M:\mathbb{R}\to\mathbb{R}$ are ridge functions and $\bm\theta_1,\dots,\bm\theta_M$ are associated projection directions (or simply ``directions''), which are constrained to lie on the unit $p$-dimensional hypersphere $\mathcal{S}^p$ for identifiability reasons. The form of the ridge functions is entirely up to the modeler, but we have seen that the chosen form can make a substantial difference in predictive performance. Importantly, nonlinear ridge functions effectively allow for interactions to enter the model. For example, consider ridge function $g(u) = u^2$, two-dimensional direction $\bm\theta = (1/\sqrt{2},1/\sqrt{2})^\prime$, and feature vector $\bm x = \left(x_1,x_2\right)^\prime$ so that $g(\bm x^\prime\bm\theta) = 0.5x_1^2 + 0.5x_2^2 + x_1x_2$, which clearly incorporates an interaction term. In their original paper, \cite{ppr} propose nonparametric smoother functions with bandwidths chosen via leave-one-out cross-validation, which are the default in R. Other types of ridge functions have since been used, including Gaussian processes \citep{gp_ppr}.

\cite{ppr} use a sequential back-fitting algorithm to optimize each direction and ridge function. The algorithm begins by optimizing one ridge function and its direction simultaneously, then adds a second ridge function and direction to fit the residuals of the first. This process is repeated until the reduction in residual variance drops below a pre-specified threshold (typically chosen via cross-validation), and $M$ is set to the total number of ridge functions at that point. We evaluate PPR as part of our bake-off in \cref{sec:app}, and find that it compares favorably to its likelihood-free competitors, but is typically out-performed by its Bayesian counterpart.


\subsection{Single Index Models}\label{sec:sim}

The Single Index Model is a single-ridge-function version of PPR, i.e., $f(\bm x) = g(\bm x^\prime \bm\theta)$ (see \cite{sim1,sim2,friedmanandtukey}). The first Bayesian version of SIM was designed by \cite{antoniadis} using a basis expansion to represent $g$, so that their model may be written
\begin{equation}\label{eq:sim}
    f(\bm x) = \bm \beta^\prime \bm b(\bm x^\prime \bm\theta | \bm t),
\end{equation}
\noindent where $\bm \beta\in\mathbb{R}^K$ is the coefficient vector of a $K$-dimensional basis expansion $\bm b: \mathbb{R}\to \mathbb{R}^K$, which is characterized by a vector of knot points $\bm t$. They use a cubic B-splines expansion for $\bm b$, so that the ridge function is piecewise with cubic polynomials between knot points and outside of knot boundaries, and the entire function is constrained to have continuous second-order derivative across $\mathbb{R}$. They set $\bm b$ \textit{a priori} with a fixed number of knot points, and knots located at fixed quantiles of $\bm x^\prime \bm\theta$. They then use a penalized least squares estimator as the prior mean of $\bm \beta$, and a Fisher-von Mises distribution (FvM), centered at a data-dependent value, as a prior for $\bm\theta\in\mathcal{S}^p$. Updates of $\bm\theta$ involve a symmetric proposal from an FvM, which requires a costly rejection sampling algorithm.

Later, \cite{wang} implemented a fully Bayesian version of SIM. They too use B-spline for the ridge function, so their model has the same form as \cref{eq:sim}, but they use a non-empirical prior for $\bm \beta$ and treat the number and location of knot points as free parameters. They also use a uniform distribution on the unit \textit{half}-sphere (for identifiability reasons) as a prior for $\bm\theta$ instead of an FvM. In addition, they perform variable selection by learning which elements $\theta_j$ of $\bm\theta$ should be zero because the corresponding input dimension is unimportant to the response surface. Metropolis-Hastings (MH) proposals are performed two elements at a time, using a clever trick involving a univariate Gaussian draw to avoid FvM entirely. Their use of fully Bayesian techniques plays a large role in paving the way for BPPR.

Finally, two Bayesian versions of SIM represent the ridge function $g$ as a GP. First, \cite{choi} provided a straightforward implementation using FvM prior and proposal distributions for $\bm\theta$. More recently, \cite{gramacy} observed that this reduces to a certain GP regression model, which reduces computation time. We include the latter implementation in our bake-off, and find that BPPR is more accurate in most cases due to the flexibility provided by additional ridge functions.

\subsection{Bayesian Projection Pursuit Regression}

We now present our novel Bayesian Projection Pursuit Regression model. As in traditional PPR, we use \cref{eq:ppr} to model $f$  with ridge functions $g_1,\dots,g_M$ in directions $\bm\theta_1,\dots,\bm\theta_M$, where $\bm\theta_m \in\mathcal{S}^p$ for $m=1,\dots,M$. In contrast to PPR, we treat the number of ridge functions $M$ as a parameter having prior distribution
\begin{equation}\label{eq:M}
    M \sim \text{Poisson}(\lambda),
\end{equation}
with $\lambda = 10$ the suggested default. In the remainder of this subsection, we recommend a form for the ridge functions, and prior distributions for the associated parameters.

\subsubsection*{Ridge Functions}

\begin{figure}[h]
    \centering
    \includegraphics[width=4.5in]{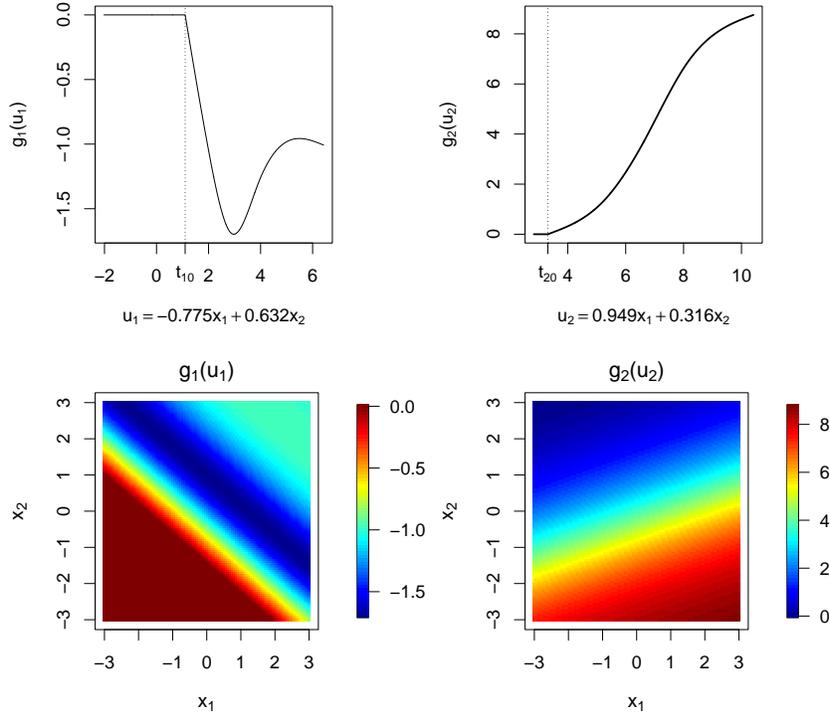}
    \caption{Two ridge function examples, plotted against one-dimensional projections (top) and on the two-dimensional input space (bottom).}
    \label{fig:ridge}
\end{figure}

To accommodate large datasets and speed computation, we use a basis expansion $\bm b_m:\mathbb{R} \to \mathbb{R}^K$ with coefficient vector $\bm \beta_m\in\mathbb{R}^K$ to represent ridge function $g_m$. We also insert an intercept term $\beta_0$, and model $f$ as
\begin{equation}\label{eq:bppr}
    f(\bm x) = \beta_0 + \sum_{m=1}^M \bm \beta_m^\prime \bm b_m(\bm x^\prime \bm\theta_m| \bm t_m),
\end{equation}
\noindent where $\bm b_m$ is a modified natural spline expansion characterized by a vector of knot points $\bm t_m = (t_{m0},\dots,t_{m(K+1)})$. Natural splines are identical to B-splines (see \cref{sec:sim}) except they are linear outside the knot boundaries, which we have found to be critical for robust prediction. We slightly modify the standard natural cubic spline expansion (see, e.g., \cite{esl}) and write the basis $\bm b_m(\cdot|\bm t_m) = \left(b_{m1}(\cdot|\bm t_m),\dots,b_{mK}(\cdot|\bm t_m)\right)^\prime$ as
\begin{equation}\label{eq:spline_expansion}
    \begin{aligned}
        b_{m1}(u|\bm t_m) &= (u - t_{m0})_+ \\
        b_{m\ell}(u|\bm t_m) &= d_{m(\ell-1)}(u|\bm t_m) - d_{mK}(u|\bm t_m), \ \ell = 2,\ldots,K \\
        d_{m\ell}(u|\bm t_m) &= \frac{\left(u - t_{m\ell}\right)_+^3 - \left(u - t_{m(K+1)}\right)_+^3}{t_{m(K+1)} - t_{m\ell}}, \ \ell = 1,\dots,K
    \end{aligned}
\end{equation}
\noindent where $d_{m\ell}(u | \bm t)$ is called the $\ell^{th}$ divided difference function and $(x)_+$ denotes the positive part of $x$. For out-of-the-box implementation, we recommend $K = 4$ basis functions as a default, implying $K+2=6$ total knot points, which we use for all examples in \cref{sec:app}. The only distinction between \cref{eq:spline_expansion} and standard natural splines is that $b_{m1}(u | \bm t) = u$ in the latter, which eliminates the need for $t_{m0}$. By using the modification in \cref{eq:spline_expansion}, we obtain $\bm b_m(u | \bm t_m) = 0$ for $u \le t_{m0}$ so that the ridge function is zero over part of the domain. Thus, larger values of $t_{m0}$ facilitate more localized prediction, while smaller values allow ridge functions to have a more global effect. \Cref{fig:ridge} shows two examples of ridge functions of this form, plotted on the projection space (top) and across the two-dimensional input space (bottom). Note that these functions have continuous second-order derivatives everywhere except at $t_{m0}$, where they are continuous but not differentiable. The power of BPPR comes from summing $M\ge 1$ such ridge functions to approximate complex response functions.

The spline representation in \cref{eq:spline_expansion} requires specification of knot points $t_{m0},\dots,$ $t_{m(K+1)}$ for each ridge function. We treat the initial knot $t_{m0}$ as a free parameter with continuous uniform prior
\begin{equation}\label{eq:eta_0}
    t_{m0} | \bm X\bm\theta_m \overset{\text{ind}}\sim \text{Unif}\left(L_m, U_m\right), \ m=1,\dots,M,
\end{equation}
where $L_m < U_m$ are carefully chosen bounds that depend on the observed projections $\bm X\bm\theta_m$. We fix the bounds according to the equations
\begin{equation}\label{eq:UL}
    \begin{aligned}
        U_m &= Q_q\left(\bm X\bm\theta_m\right), \\
        L_m &= U_m - \left(U_m - Q_0\left(\bm X\bm\theta_m\right)\right)/p_0, 
    \end{aligned}
\end{equation}
where $Q_q\left(\bm X\bm\theta_m\right)$ is the $q^{th}$ quantile of the observed projections. We recommend choosing $q$ such that $g(\bm x_i^\prime\bm\theta_m)$ is nonzero for at least $\sim 20$ observed inputs $\bm x_i$ to avoid overfitting at the boundary of the input space. The lower bound $L_m$ is set to achieve a good mix between localized and global ridge functions, with user-chosen $p_0\in[0, 1]$ interpreted as the prior probability that $t_{m0} < Q_0\left(\bm X\bm\theta_m\right)$; i.e. $p_0$ is the prior probability that $g(\bm x_i^\prime\bm\theta_m) = 0$ for at least one $\bm x_i$. We recommend using $p_0 = 2/3$ as a default. Finally, conditional on $t_{m0}$, we fix the remaining knots at equally-spaced quantiles of the observed projections that are greater than $t_{m0}$, i.e.,
\begin{equation}\label{eq:eta_-0}
    t_{m\ell} = Q_{(\ell-1)/K}\left(\{\bm x_i^\prime\bm\theta_m : \bm x_i^\prime\bm\theta_m > t_{m0}\}\right), \ \ell = 1,\dots,K+1,
\end{equation}
where $Q_{(\ell-1)/K}\left(\{u_i\}\right)$ is the $(\ell-1)/K$ sample quantile of $\{u_i\}_{i=1}^n$.

\subsubsection*{Variable Selection and Coefficient Estimation}

To facilitate variable selection, we introduce auxiliary variables $a_1,\dots,a_M$ and $\mathcal{J}_1,\dots,$ $\mathcal{J}_M$, where $a_m\in \{1,\dots,p\}$ is the number of active dimensions, i.e., the number of nonzero elements in $\bm\theta_m$, and $\mathcal{J}_m\subseteq\{1,\dots,p\}$ is a set of size $|\mathcal{J}_m| = a_m$ containing the indices of the active dimensions. For the $a_m$, we use independent discrete uniform priors with support on $\{1,\dots,A\}$, where $A$ is the user-specified maximum number of active dimensions for any single ridge function. We typically set $A = \min\{3, p\}$, which effectively allows for interactions up to order three. Given the number of active dimensions $a_m$, we also use independent discrete uniform priors
\begin{equation}\label{eq:J}
    \pi(\mathcal{J}_m|a_m) =
        \begin{cases}
            {{p}\choose{a_m}}^{-1} & \text{ if } |\mathcal{J}_m| = a_m\\
            0 & \text{ otherwise}
        \end{cases}, \ m=1,\dots,M,
\end{equation}
\noindent for the active dimension indices, so that equal weight is placed on each combination of indices. Other priors might be desired if some inputs or interactions are known \textit{a priori} to have a more substantial association with the response.

Given $a_m$ and $\mathcal{J}_m$, we assign the $\bm\theta_m$ independent uniform priors on the unit sphere, with the inactive dimensions constrained to be zero. Writing $\bm\theta_{\mathcal{J}_m}$ for the active (nonzero) elements of $\bm\theta_m$ and $\bm\theta_{-\mathcal{J}_m}$ for the inactive (zero) elements, we have
\begin{equation}\label{eq:theta}
    \bm\theta_m | \left(a_m, \mathcal{J}_m\right) \overset{\text{ind}}\sim \text{Unif}\left(\{\bm\theta
    \in\mathcal{S}^p: \bm\theta_{-\mathcal{J}_m} = \bm 0_{p-a_m}\}\right), \ m=1,\dots,M,
\end{equation} 
where $\bm 0_{p-a_m}$ is a vector of $p-a_m$ zeroes. We note here that it is important to standardize inputs to have mean zero and variance one. This, combined with the uniform prior, assures that all input features have equal opportunity to contribute to the projection. More complicated priors might be used, but a uniform prior simplifies elicitation and facilitates out-of-the-box implementation.

Finally, we use the Zellner-Siow prior \citep{zellner-siow} for the combined regression coefficient vector $\bm \beta = (\beta_0,\bm \beta_1^\prime,\dots,\bm \beta_M^\prime)^\prime$ and the response variance $\sigma^2$. Letting $\bm B_m = \left[\bm b_m(\bm x_1^\prime \bm\theta_m | \bm t_m) \ \dots \ \bm b_m(\bm x_n^\prime \bm\theta_m | \bm t_m)\right]^\prime$ be the $n\times K$ basis matrix for the $m^{th}$ ridge function, and $\bm B = \left[\bm 1_n \ \bm B_1 \ \dots \ \bm B_M\right]\in\mathbb{R}^{n}\times \mathbb{R}^{1 + KM}$ the concatenation of an intercept vector $\bm 1_n$ with the basis matrices from all ridge functions, the Zellner-Siow prior is
\begin{equation}\label{eq:zellner-siow}
    \begin{aligned}
        &\bm \beta \:|\: \left(\bm B, \sigma^2, \tau\right) \sim \mathcal{N}_{1 + KM}\left(\bm 0_{1 + KM}, \tau\sigma^2 \left(\bm B^\prime \bm B\right)^{-1}\right)\\
        &\pi(\sigma^2) \propto 1/\sigma^2\\
        &\tau \sim \text{Inv-Gamma}(1/2, n/2).
    \end{aligned}
\end{equation}
We have experimented with several other priors for $\bm \beta$, but \cref{eq:zellner-siow} has the benefit of being non-informative, while at the same time avoiding model selection paradoxes associated with many non-informative priors (see \cite{mix_g-prior}). It performs well in a variety of settings, as illustrated in \cref{sec:app}.

\subsection{Posterior Sampling}

Let $\bm\phi = (\bm\phi_1,\dots,\bm\phi_M)$ be the vector of parameters needed to construct basis matrix $\bm B$, where $\bm\phi_m = (a_m, \mathcal{J}_m, \bm\theta_m, t_{m0})$ are the parameters relevant to $\bm B_m$ for $m=1,\dots,M$. Given $t_{m0}$, the remaining knots are fixed according to \cref{eq:eta_-0}, so they need not be included in $\bm \phi_m$. To obtain valid estimates and uncertainty quantification for all model parameters, we sample from their joint posterior distribution
\begin{equation}\label{eq:posterior}
    \pi(M, \bm \phi, \bm\beta, \sigma^2, \tau | (\bm X, \bm y))\propto \pi(\bm y | \bm B, \bm \beta, \sigma^2) \pi(\bm\beta, \sigma^2, \tau | \bm B) \ \pi(M) \prod_{m=1}^M \pi(\bm\phi_m),
\end{equation}
\noindent with $\pi(M)$ given in \cref{eq:M}, $\pi(\bm\phi_m) = \pi(a_m)\pi(\mathcal{J}_m | a_m)\pi(\bm\theta_m | a_m, \mathcal{J}_m)\pi(t_{m0} | \bm X\bm\theta_m)$ defined by eqs. (\ref{eq:eta_0})-(\ref{eq:theta}), and $\pi(\bm\beta, \sigma^2, \tau | \bm B)$ the Zellner-Siow prior in \cref{eq:zellner-siow}.

For $s = 0,\dots,N_{\text{mcmc}}$, let $\bm\phi^{(s)} = (\bm\phi_1^{(s)},\dots,\bm\phi_{M^{(s)}}^{(s)})$ represent the basis parameters of posterior sample $s$, with analogous notation for the remaining parameters. To begin our MH algorithm, we initialize $\bm\phi^{(0)}$ with $M^{(0)} = 0$. Then, for $s = 1,\dots,N_{\text{mcmc}}$, we sample $\left(M^{(s)},\bm \phi^{(s)}\right)$ from the posterior distribution by performing either a birth, death, or change step, where the type of step is chosen uniformly at random. A birth step proposes to add a new ridge function by augmenting $\bm\phi$, while a death step proposes to delete a randomly chosen ridge function $g_m$, along with its corresponding parameters $\bm\phi_m$. Both of these require RJMCMC techniques, but are simplified by marginalizing $(\bm \beta, \sigma^2)$ out of the likelihood. A change step selects an index $m\in\{1,\dots,M\}$ and proposes to modify $\bm\phi_m$. After performing one of these three steps, the remaining parameters $\bm\beta$, $\sigma^2$ and $\tau$ are drawn from their full conditional distributions via Gibbs steps. The overall MCMC algorithm is given in \cref{alg:alg1}, with birth, death, and change proposals specified more completely in the following.
\begin{algorithm}[H]
\caption{MCMC}
\label{alg:alg1}
\begin{algorithmic}[1]
\onehalfspacing
\State \texttt{Sample} $T\sim\text{Unif}(\{\text{b},\text{d},\text{c}\})$
\If{$T = \text{b}$}
\State \texttt{Sample} $\left(M^{(s)},\bm\phi^{(s)}\right)$ via a birth proposal \Comment{see \cref{alg:alg2}}
\ElsIf{$T = \text{d}$}
\State \texttt{Sample} $\left(M^{(s)},\bm\phi^{(s)}\right)$ via a death proposal \Comment{see \cref{alg:alg3}}
\Else
\State \texttt{Set} $M^{(s)} \leftarrow M^{(s-1)}$
\State \texttt{Sample} $\bm\phi^{(s)}$ via a change proposal \Comment{see \cref{alg:alg4}}
\EndIf
\For{$m=1,\dots,M$}
\State \texttt{Set} $\bm B_m^{(s)} \leftarrow \left[\bm b_m(\bm x_1^\prime \bm\theta_m^{(s)} | \bm t_m^{(s)}) \ \dots \ b_m(\bm x_n^\prime \bm\theta_m^{(s)} | \bm t_m^{(s)})\right]^\prime$
\EndFor
\State \texttt{Set} $\bm B^{(s)} \leftarrow \left[\bm 1_n \ \bm B_1^{(s)} \ \dots \ \bm B_M^{(s)}\right]$
\State \texttt{Set} $\bm\Lambda^{(s)} \leftarrow \frac{\tau^{(s-1)}}{1 + \tau^{(s-1)}}\left({\bm B^{(s)}}^\prime\bm B^{(s)}\right)^{-1}$
\State \texttt{Sample} $\bm\beta^{(s)} \sim \mathcal{N}_{1 + KM^{(s)}}\left(\bm\Lambda^{(s)}{{\bm B^{(s)}}^\prime}\bm y, \ {\sigma^2}^{(s-1)}\bm\Lambda^{(s)}\right)$ \Comment{Gibbs step}
\State \texttt{Sample} ${\sigma^2}^{(s)} \sim \text{Inv-Gamma}\left(\frac{n}{2}, \ \frac{||\bm y - \bm B^{(s)}\bm\beta^{(s)}||^2}{2}\right)$ \Comment{Gibbs step}
\State \texttt{Sample} ${\tau}^{(s)} \sim \text{Inv-Gamma}\left(1 + \frac{KM}{2}, \ \frac{n + ||\bm B^{(s)}\bm\beta^{(s)}||^2/{{\sigma^2}^{(s)}}}{2}\right)$ \Comment{Gibbs step}
\end{algorithmic}
\end{algorithm}

\subsubsection*{Birth Proposal}

A birth step generates a proposal $\bm\phi^\star = \left(a^\star, \mathcal{J}^\star, \bm\theta^\star, t_0^\star\right)$ and uses the MH algorithm to either accept $\bm\phi^\star$ and insert it at a random location $m^\star\in\{1,\dots,M^{(s-1)}+1\}$ in $\bm\phi^{(s)}$ so that $\bm\phi^{(s)} = (\bm\phi^{(s-1)}_1,\dots,\bm\phi^{(s-1)}_{m^\star-1},\bm\phi^\star, \bm\phi^{(s-1)}_{m^\star},\dots,\bm\phi^{(s-1)}_{M^{(s-1)}})$ and $M^{(s)}=M^{(s-1)} + 1$, or retain $\bm\phi^{(s)}=\bm\phi^{(s-1)}$ and $M^{(s)}=M^{(s-1)}$. Details are given in \cref{alg:alg2}. The proposal distributions for $\bm\theta^\star$ and $t_0^\star$ are identical to their prior distributions, while the number of active dimensions $a^\star$ and the set of active indices $\mathcal{J}^\star$ are proposed using an adaptive strategy, modified from its implementation in the BMARS model \citep{nottkukduc}.
\begin{algorithm}[H]
\caption{Birth Proposal}
\label{alg:alg2}
\begin{algorithmic}[1]
\onehalfspacing
\State \texttt{Sample} $a^\star \sim \text{Categorical}\left(\{1, \ldots A\},\: w_1^{(s-1)}, \ldots w_A^{(s-1)}\right)$ \Comment{see \cref{eq:a_weights}}
\If{$a^\star = 1$}
\State \texttt{Sample} $\mathcal J^\star|a^\star \sim \text{Unif}\left(\{1, \ldots p\}\right)$
\Else
\State \texttt{Sample} $\mathcal J^\star | a^\star \sim \mathcal W_{a^\star}\left(\upsilon_1^{(s-1)}, \ldots \upsilon_p^{(s-1)}\right)$ \Comment{see \cref{eq:J_weights}}
\EndIf
\State \texttt{Sample} $\bm\theta^\star | \left(a^\star, \mathcal J^\star\right) \sim \text{Unif}\left(\{\bm\theta
    \in\mathcal{S}^p: \bm\theta_{-\mathcal{J}^\star} = \bm 0_{p-a^\star}\}\right)$ \Comment{prior distribution; \cref{eq:theta}}
\State \texttt{Set} $U^\star \leftarrow Q_q\left(\bm X\bm\theta^\star\right)$ \Comment{\cref{eq:UL}}
\State \texttt{Set} $L^\star \leftarrow U^\star - \left(U^\star - Q_0\left(\bm X\bm\theta^\star\right)\right)/p_0$ \Comment{\cref{eq:UL}}
\State \texttt{Sample} $t^\star_0 | \bm X\bm\theta^\star \sim \text{Unif}\left(L^\star, U^\star\right)$ \Comment{prior distribution; \cref{eq:eta_0}}
\State \texttt{Set} $\bm\phi^\star \leftarrow \left(a^\star, \mathcal{J}^\star, \bm\theta^\star, t_0^\star\right)$
\State \texttt{Sample} $m^\star \sim \text{Unif}\left(\{1,\dots,M^{(s-1)}+1\}\right)$ \Comment{proposed index of $\bm\phi^\star$ in $\bm\phi^{(s)}$}
\State \texttt{Set} $p_\text{acc}^{(b)} \leftarrow \min\{1, \alpha^{(b)}\}$ \Comment{see \cref{eq:mhb}}
\State \texttt{Sample} $V \sim \text{Unif}(0, 1)$
\If{$V < p_{\text{acc}}^{(\text{b})}$} 
\State \texttt{Set} $\bm\phi^{(s)} \leftarrow \left(\bm\phi^{(s-1)}_1,\dots,\bm\phi^{(s-1)}_{m^\star-1}, \bm\phi^\star,\bm\phi^{(s-1)}_{m^\star},\dots,\bm\phi^{(s-1)}_{M^{(s-1)}}\right)$
\State \texttt{Set} $M^{(s)} \leftarrow M^{(s-1)} + 1$
\Else
\State \texttt{Set} $\bm\phi^{(s)} \leftarrow \bm\phi^{(s-1)}$
\State \texttt{Set} $M^{(s)} \leftarrow M^{(s-1)}$
\EndIf
\end{algorithmic}
\end{algorithm}

The idea behind the adaptive sampling strategy is to propose numbers $a^\star$ and indices $\mathcal{J}^\star$ of features that are more frequently used in the current set of ridge functions. At iteration $s$, $a^\star$ is drawn randomly from the set $\{1,\dots,A\}$ with sampling weights 
\begin{equation}\label{eq:a_weights}
    \omega_a^{(s-1)} \propto \omega_0 + \sum_{m=1}^{M^{(s-1)}} \bm 1\left(a_m^{(s-1)} = a\right), a \in \{1,\ldots A\}.
\end{equation} 
for $j = 1,\dots,p$. The weights are proportional to the number of ridge functions currently using exactly $a$ features, plus a user-specified constant $\omega_0$. Setting $\omega_0$ large leads to an approximately uniform proposal for $a^\star$, while $\omega_0$ small encourages selection of values that are frequently used in the current model. If $a^\star = 1$, the single index of $\mathcal{J}^\star$ is sampled uniformly from the set $\{1, \ldots, p\}$. However, if $a^\star \geq 2$ then we sample $a^\star$ distinct values without replacement from $\{1,\ldots p\}$, with sampling weights
\begin{equation}\label{eq:J_weights}
\upsilon_j^{(s-1)} \propto \upsilon_0 + \sum_{m=1}^{M^{(s-1)}} \bm 1(j \in \mathcal{J}_m^{(s-1)}), \ j=1,\dots,p.
\end{equation}
Similar to before, these weights are proportional to the number of times feature $j$ is used in the current set of ridge functions, plus a user-specified constant $\upsilon_0$. This is equivalent to sampling $\mathcal J^\star|a^\star \sim \mathcal{W}_{a^\star}\left(\upsilon_1^{(s-1)}, \ldots \upsilon_p^{(s-1)}\right)$ where $\mathcal W_n(w_1, \ldots w_N)$ denotes Wallenius’ noncentral hypergeometric distribution \citep{wallenius}. We set $\omega_0 = \upsilon_0 = 1$ for all examples in \cref{sec:app}, but larger values may be desirable in some cases. This adaptive proposal improves posterior mixing, especially when some of the input dimensions are inert. For more details and examples of its implementation, see \cite{nottkukduc} and \cite{bass}.

Letting $(M, \bm B, \tau, \omega_{a^\star}, \upsilon_j) =  (M^{(s-1)}, \bm B^{(s-1)}, \tau^{(s-1)}, \omega_{a^\star}^{(s-1)}, \upsilon_j^{(s-1)})$ represent values from the previous iteration, the MH ratio for a birth is
\begin{equation}
\label{eq:mhb}
\begin{aligned}
    \alpha^{(b)} = &\left[\frac{\pi\left(\bm y | \bm B^\star, \tau\right)}{\pi\left(\bm y | \bm B, \tau\right)}\right] \times \left[\frac{\pi(M^\star)\pi(a^\star)\pi(\mathcal{J}^\star | a^\star)\pi(\bm\theta^\star | \mathcal{J}^\star, a^\star)\pi(t_0^\star | \bm X\bm\theta^\star)}{\pi(M)}\right]\\
    &\times \left[\frac{1/M^\star}{\omega_{a^\star}\times\mathcal{W}_{a^\star}\left(\mathcal{J}^\star | \upsilon_1,\dots,\upsilon_p\right)\pi(\bm\theta^\star | \mathcal{J}^\star, a^\star)\pi(t_0^\star | \bm X\bm\theta^\star)/M^\star}\right]\\
    = &\left[\left(1 + \tau\right)^{-\frac{K}{2}}\left(\frac{\bm y^\prime \bm y - \frac{\tau}{1 + \tau} \bm y^\prime \bm B^\star \left({\bm B^\star}^\prime \bm B^\star\right)^{-1} {\bm B^\star}^\prime \bm y}{\bm y^\prime \bm y - \frac{\tau}{1 + \tau} \bm y^\prime \bm B \left({\bm B}^\prime \bm B\right)^{-1} {\bm B}^\prime \bm y}\right)^{-n/2}\right]\\
    &\times \left[\frac{\lambda / \left(M^\star A{p \choose a^\star}\right)}{\omega_{a^\star}\times  \mathcal{W}_{a^\star}\left(\mathcal{J}^\star \big| \upsilon_1,\dots,\upsilon_p\right)}\right].
\end{aligned}
\end{equation}
\noindent In the first expression, the first bracketed term is the ratio of marginal likelihoods, integrating out $\bm \beta$ and $\sigma^2$. The second term is the ratio of prior distributions, where we exclude terms common to the proposed and current models due to cancellation, and the third is the ratio of proposal distributions. The second expression reduces $\alpha^{(b)}$ to two terms: the expanded ratio of marginal likelihoods and a simplified expansion of the prior and proposal ratios.

\subsubsection*{Death Proposal}

A death step randomly selects an index $m^\star\in\{1,\dots,M^{(s-1)}\}$ of a ridge function to delete, and uses the MH algorithm to either accept the deletion and set $\bm\phi^{(s)}=\bm\phi^{(s-1)}_{-\{m^\star\}}$ and $M^{(s)}=M^{(s-1)} - 1$, or retain $\bm\phi^{(s)}=\bm\phi^{(s-1)}$ and $M^{(s)}=M^{(s-1)}$. The death proposal is outlined in \cref{alg:alg3}, where the MH ratio $\alpha^{(d)}$ is the inverse of the birth ratio, and is given explicitly in the supplemental materials.
\begin{algorithm}[H]
\caption{Death Proposal}
\label{alg:alg3}
\begin{algorithmic}[1]
\onehalfspacing
\State \texttt{Sample} $m^\star \sim \text{Unif}\left(\{1,\dots,M^{(s-1)}\}\right)$
\State \texttt{Set} $p_\text{acc}^{(d)} \leftarrow \min\{1, \alpha^{(d)}\}$ \Comment{see supplemental materials}
\State \texttt{Sample} $V \sim \text{Unif}(0, 1)$
\If{$V < p_{\text{acc}}^{(\text{d})}$} 
\State \texttt{Set} $\bm\phi^{(s)} \leftarrow \bm\phi^{(s-1)}_{-\{m^\star\}}$
\State \texttt{Set} $M^{(s)} \leftarrow M^{(s-1)} - 1$
\Else
\State \texttt{Set} $\bm\phi^{(s)} \leftarrow \bm\phi^{(s-1)}$
\State \texttt{Set} $M^{(s)} \leftarrow M^{(s-1)}$
\EndIf
\end{algorithmic}
\end{algorithm}

\subsubsection*{Change Proposal}

Finally, a change step randomly selects an index $m^\star\in\{1,\dots,M^{(s-1)}\}$ of a ridge function and proposes a modification to $\bm\phi_{m^\star}^{(s-1)}$. The proposal leaves the number $a_{m^\star}^{(s-1)}$ and indices $\mathcal{J}_{m^\star}^{(s-1)}$ of active dimensions unchanged, but updates the projection direction $\bm\theta_{m^\star}^{(s-1)}$ to $\bm\theta^\star$---with the active dimensions $\bm\theta^\star_{\mathcal{J}_m^{(s-1)}}$ drawn from a proposal distribution $\mathcal{P}(\cdot)$---and updates the knot point $t_{m^\star 0}^{(s-1)}$ to $t_0^\star$ drawn from its prior distribution. The change proposal is outlined in \cref{alg:alg4}, with the MH ratio $\alpha^{(c)}$ given in the supplemental materials.
\begin{algorithm}[H]
\caption{Change Proposal}
\label{alg:alg4}
\begin{algorithmic}[1]
\onehalfspacing
\State \texttt{Sample} $m^\star \sim \text{Unif}\left(\{1,\dots,M^{(s-1)}\}\right)$
\State \texttt{Sample} $\bm\theta^\star_{\mathcal{J}_{m^\star}^{(s-1)}} \big| \left(a_{m^\star}^{(s-1)}, \mathcal{J}_{m^\star}^{(s-1)}, \mu, \kappa\right) \sim \mathcal{P}(\cdot | \mu, \kappa)$ \Comment{see \cref{eq:powersphere}}
\State \texttt{Set} $\bm\theta^\star_{-\mathcal{J}_{m^\star}^{(s-1)}} \leftarrow 0$ \Comment{\cref{eq:theta}}
\State \texttt{Set} $U^\star \leftarrow \{\bm X \bm\theta^\star\}_{(Q)}$ \Comment{\cref{eq:UL}}
\State \texttt{Set} $L^\star \leftarrow \frac{1}{q}\{\bm X \bm\theta^\star\}_{(0)} - \frac{1-q}{q}U^\star$ \Comment{\cref{eq:UL}}
\State \texttt{Sample} $t^\star_0 | \left(\bm X,  \bm\theta^\star\right) \sim \text{Unif}\left(L^\star, U^\star\right)$ \Comment{prior distribution; \cref{eq:eta_0}}
\State \texttt{Set} $\bm\phi^\star \leftarrow \left(a_{m^\star}^{(s-1)}, \mathcal{J}_{m^\star}^{(s-1)}, \bm\theta^\star, t_0^\star\right)$
\State \texttt{Set} $p_\text{acc}^{(c)} \leftarrow \min\{1, \alpha^{(c)}\}$ \Comment{see supplemental materials}
\State \texttt{Sample} $V \sim \text{Unif}(0, 1)$
\If{$V < p_{\text{acc}}^{(\text{c})}$} 
\State \texttt{Set} $\bm\phi^{(s)} \leftarrow \left(\bm\phi^{(s-1)}_1,\dots,\bm\phi^{(s-1)}_{m^\star-1}, \bm\phi^\star,\bm\phi^{(s-1)}_{m^\star+1},\dots,\bm\phi^{(s-1)}_{M^{(s-1)}}\right)$
\Else
\State \texttt{Set} $\bm\phi^{(s)} \leftarrow \bm\phi^{(s-1)}$
\EndIf
\end{algorithmic}
\end{algorithm}

To generate a proposal for the active dimensions $\bm\theta^\star_{\mathcal{J}_{m^\star}^{(s-1)}}$ of the projection direction, we use the $a^\star$-dimensional power spherical distribution \citep{powersphere}:
\begin{equation}\label{eq:powersphere}
    \mathcal{P}_{a^\star}(\bm\theta | \mu, \kappa) \propto (1 + \mu^\prime\bm\theta)^{\kappa},
\end{equation}
\noindent which is symmetric around mode $\mu = \bm\theta_{\mathcal{J}_{m^\star}^{(s-1)}}^{(s-1)}$, with precision $\kappa$. We typically set $\kappa = 1000$ for good posterior mixing. Like FvM, the power spherical distribution has support on the unit hypersphere. Unlike FvM, it also comes with a closed-form CDF and is equipped with a direct sampling method that facilitates fast execution.

\subsubsection*{Bayesian Back-Fitting as an Alternative MCMC Algorithm}

Here, we consider Bayesian back-fitting as an alternative to the MCMC algorithm described above. Given a fixed number $M$ of ridge functions, Bayesian back-fitting updates the pairs $(\bm\phi_1, \bm\beta_1),\dots,$ $(\bm\phi_M, \bm\beta_M)$ one at a time, holding all other pairs fixed at each step. Specifically, for $m = 1,\dots,M$ it first performs a change step to sample $\bm\phi_m | (\bm\phi_{-\{m\}}, \bm\beta_{-\{m\}}, \tau, \sigma^2)$, then samples $\bm\beta_m$ with a Gibbs step. It then samples $\beta_0$, $\sigma^2$, and $\tau$ with Gibbs steps, and $M$ with either a birth or death step, holding all other basis parameters and coefficients fixed. The main advantage of this approach is computational: in Bayesian back-fitting, birth, death and change steps involve operations on $\bm B_m$, which is a relatively small $n\times K$ matrix. In constrast, \cref{alg:alg1} requires operations on the $n\times (KM+1)$ matrix $\bm B$, which involves significantly more computation time per step.

We implemented the Bayesian back-fitting algorithm, but found that the mixing was much slower than with \cref{alg:alg1}, so that many more birth, death, and change steps were required to obtain a comparable fit. Thus, we recommend the implementation described in \cref{alg:alg1} and we use this approach for all examples in \cref{sec:app}.

\subsection{Categorical Input}

Suppose we have $L$ categories $\mathcal{C}_1,\dots,\mathcal{C}_L$ to which a categorical input variable $c$ may belong, and let $c=\mathcal{C}_\ell$ indicate that $c$ belongs to category $\mathcal{C}_\ell$. We code $c$ as a vector of $p^{(d)} = L - 1$ ``dummy'' variables $d_1,\dots,d_{p^{(d)}}$, where $d_\ell = \bm 1(c=\mathcal{C}_\ell)$. With $p^{(\mathbb{R})}$ real-valued variables also present, this gives us a total of $p = p^{(\mathbb{R})} + p^{(d)}$ inputs, denoted $\bm x = \left(x_1,\dots,x_{p^{(\mathbb{R})}},d_1,\dots,d_{p^{(d)}}\right)^\prime \in \mathbb{R}^p$. In the case of multiple categorical variables, additional corresponding dummy variables may be appended to $\bm x$. As with real-valued inputs, we represent a ridge function as $g(\bm x) = \bm \beta^\prime \bm b(\bm x^\prime \bm\theta)$, where $\bm\theta$ has $a$ active dimensions indexed by $\mathcal{J}$, and we generally recommend using basis functions in the form of \cref{eq:spline_expansion}. The only difference is when $\mathcal{J}$ \textit{exclusively} contains indices of dummy variables: in this case, $\bm x^\prime\bm\theta$ spans only a small set of discrete values so a spline basis no longer makes sense. Instead, for an active dimension set $\mathcal{J}$ corresponding exclusively to $a$ dummy variables $d_{\ell_1},\dots,d_{\ell_a}$, we set
\begin{equation}\label{eq:cat_ridge}
g(\bm x) = \beta \left(1 - \prod_{k=1}^a(1 - d_{\ell_k})\right)
\end{equation}
\noindent so that $g(\bm x)$ becomes an indicator that $\bm x$ is in \textit{at least one} of the categories $\mathcal{C}_{\ell_1},\dots,\mathcal{C}_{\ell_a}$, multiplied by $\beta\in\mathbb{R}$. We also experimented with $g(\bm x) = \beta \prod_{k=1}^a d_{\ell_k}$ and $g(\bm x) = \beta (d_{\ell_1},\dots,d_{\ell_a})^\prime \bm\theta$, but the form in \cref{eq:cat_ridge} performs best in most scenarios.

In the presence of $p^{(\mathbb{R})}$ real-valued features and $p^{(d)}$ dummy variables, we often find it beneficial to increase the maximum number of active dimensions $A$. Accordingly, we recommend using $A = \min\left\{3, p^{(\mathbb{R})}\right\} + \min\left\{3, \lceil{p^{(d)}/2\rceil}\right\}$ as the default.

\subsection{Multivariate Response}

We use the framework of \cite{plume} to model a set of multivariate responses $\bm y_1,\dots,\bm y_n\in\mathbb{R}^D$ for $D > 1$. In this framework, we first project the responses onto an orthogonal basis matrix $\bm H = \left[\bm h_1 \ \dots \ \bm h_D\right]\in\mathbb{R}^D\times\mathbb{R}^D$ (e.g. a principle component basis) to obtain transformed responses
\begin{equation}\label{eq:y_decomp}
    \bm \eta_i = \bm H' \bm y_i\in\mathbb{R}^D; \ i = 1,\dots,n,
\end{equation}
\noindent and then model each element $\eta_{id}$ of the transformed response independently as
\begin{equation}\label{eq:mod_multivar}
    \eta_{id} \overset{\text{ind}}{\sim} \mathcal{N}(f_d(\bm x_i), \sigma_d^2), \ d = 1,\dots,D,
\end{equation}
\noindent where $\bm x_i\in\mathbb{R}^p$ is the input vector corresponding to $\bm y_i$, $f_d:\mathbb{R}^p\to\mathbb{R}$ is the mean function and $\sigma_d^2$ is the residual variance for dimension $d$ of the transformed response. The task then reduces to estimating $(f_d, \sigma_d^2)$ independently for $d = 1,\dots,D$, which may be done in parallel. Note that a response may be decomposed as $\bm y_i = \bm H \bm\eta_i$, since $\bm H^\prime = \bm H^{-1}$ for orthogonal $\bm H$. Thus, having obtained a posterior sample $\bm\eta^{(s)}_\star$ of a transformed response $\bm\eta_\star=\bm H^\prime\bm y_\star$, we construct a posterior predictive sample as $\bm y_\star^{(s)} = \bm H\bm\eta_\star^{(s)}$.

If $D$ is large or if the elements of $\bm y_i$ are sufficiently correlated, one might desire to represent $\bm y_i$ using a truncated orthogonal basis, so that $\bm H\in\mathbb{R}^D\times\mathbb{R}^{D^-}$ for $D^- < D$. The transformed response $\bm\eta_i = \bm H^\prime \bm y_i\in\mathbb{R}^{D^-}$ is obtained in the same manner, but the decomposition becomes $\bm y_i = \bm H\bm\eta_i + \bm r_i$, where $\bm r_i$ is the truncation error. Rather than trying to model $\bm r_i$, we ignore it in our current implementation under the assumption that $D^-$ is large enough to make it negligible.

\section{Applications}\label{sec:app}

In this section, we demonstrate inference, prediction, and uncertainty quantification with BPPR. We start by modeling data simulated from the Friedman function, then emulate real computer experiment data simulating storm surges during hurricanes under various input settings, and then model two simulated datasets with multivariate response. We finish by conducting a two-part bake-off between BPPR and seven other emulators, where the tasks are to model (i) responses simulated under 20 simulation scenarios, mostly from the computer model emulation literature, and (ii) 23 real datasets, some from computer and lab experiments and others from observational settings. All BPPR models are fit using our publicly available R package, with hyperparameters set to the default settings recommended in the previous section. In terms of out-of-sample prediction accuracy, BPPR is competitive in all cases, and is often the best choice. BPPR is also competitive in accurately quantifying its uncertainty.

\subsection{Friedman Function}\label{sec:friedman}

Our first dataset is generated via the Friedman function \citep{friedman_function}:
\begin{equation*}\label{eq:friedman}
    f(\bm x) = 10\sin(\pi x_1 x_2) + 20(x_3 - 0.5)^2 + 10 x_4 + 5 x_5 + 0 x_6,
\end{equation*}
\noindent which is a function of $p = 6$ variables $\bm x = (x_1,\dots,x_6)$, the last of which is inert. The function is a sum of a nonlinear interaction between $x_1$ and $x_2$, a nonlinear function of $x_3$, and a linear combination of $x_4$ and $x_5$.

We simulate $n = 300$ input-response pairs $(\bm x_i, y_i)$, with $\bm x_i \overset{\text{i.i.d.}}\sim\text{Unif}([0, 1]^p)$, and $y_i\overset{\text{ind}}\sim\mathcal{N}(f(\bm x_i), 1)$ for $i = 1,\dots,n$. We then fit a BPPR model with all default hyperparameters as described above. The sampling algorithm is run for 20,000 iterations, which takes about 24 seconds on a standard laptop. Convergence is then verified via a traceplot, effective sample size, and split $\hat{R}$ \citep{rhat} calculations for the posterior chain of $\sigma$ (see the supplemental materials for details) and the final 2,000 draws are used for posterior inference. Posterior point estimates and 95\% prediction intervals for the noisy response $\bm y$ are then calculated, along with 95\% credible intervals for the noiseless response function. We find that the observed responses are contained within their prediction intervals 97.7\% of the time, and that the Friedman function is contained in 90.0\% of its credible intervals at the sample locations. The posterior mean of $\sigma$ is estimated to be 1.05, with a 95\% credible interval ranging from 0.97 to 1.14, indicating an adequate fit of the true data-generating process.

\begin{figure}
    \centering
    \includegraphics[width = 5in, keepaspectratio]{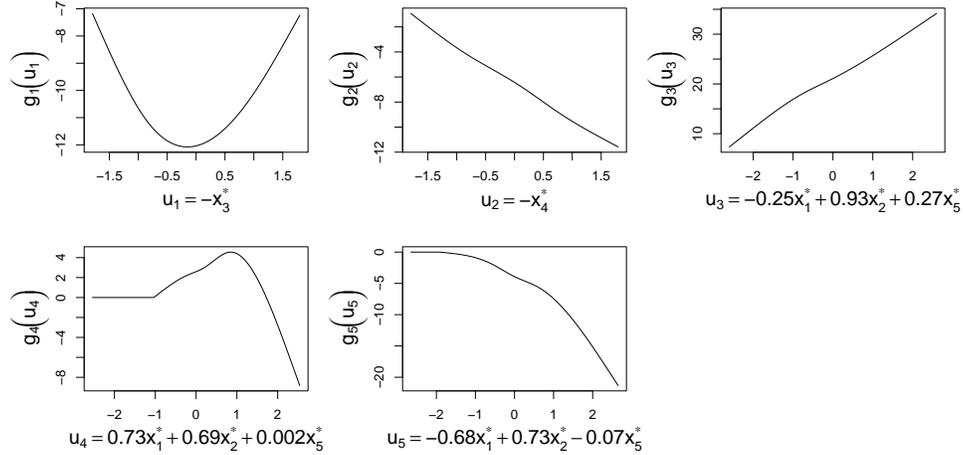}
    \caption{Ridge functions from a single posterior sample of the BPPR model of the Friedman function, where $x^\star_j$ is the standardized version of input feature $x_j$ with zero mean and unit variance.}
    \label{fig:friedman_ridge}
\end{figure}

For all post-burn-in iterations, the MCMC algorithm settles on $M = 5$ ridge functions, corresponding to active index sets $\mathcal{J}_1 = \{3\}$, $\mathcal{J}_2 = \{4\}$, and $\mathcal{J}_3 = \mathcal{J}_4 = \mathcal{J}_5 = \{1,2,5\}$. The first and second ridge functions explain the additive effects of $x_3$ and $x_4$, while the remaining ridge functions together account for the interaction between $x_1$ and $x_2$ and the additive effect of $x_5$. Ridge functions from a single posterior sample are plotted in \cref{fig:friedman_ridge}, with projection formulas printed on the horizontal axis.

\begin{figure}[h]
    \centering
    \includegraphics[width = 4.5in, keepaspectratio]{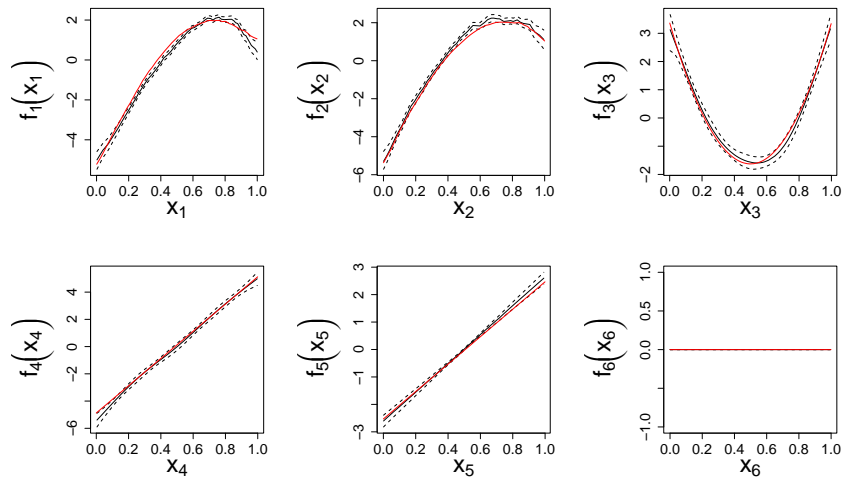}
    \caption{One-way ALE plots for the BPPR model of the Friedman function (black), with the ALE plots for the actual Friedman function (red).}
    \label{fig:friedman_ALE}
\end{figure}

Accumulated Local Effects plots (ALE plots) are a useful, recently developed tool for visualizing one- and two-way marginal effects of features on the response function (see \cite{ALE}). \Cref{fig:friedman_ALE} displays one-way posterior mean ALE plots for each of the six predictors, with red lines showing the actual marginal effects corresponding to the Friedman function and dotted lines showing 95\% credible interval bands. Note that the estimated marginal effects are close to the truth for all features, and that the estimated marginal effect of $x_6$ is rightly zero, since $6\not\in\mathcal{J}_m$ for any $\mathcal{J}_m$.

\begin{figure}
    \centering
    \includegraphics[width = 4in, keepaspectratio]{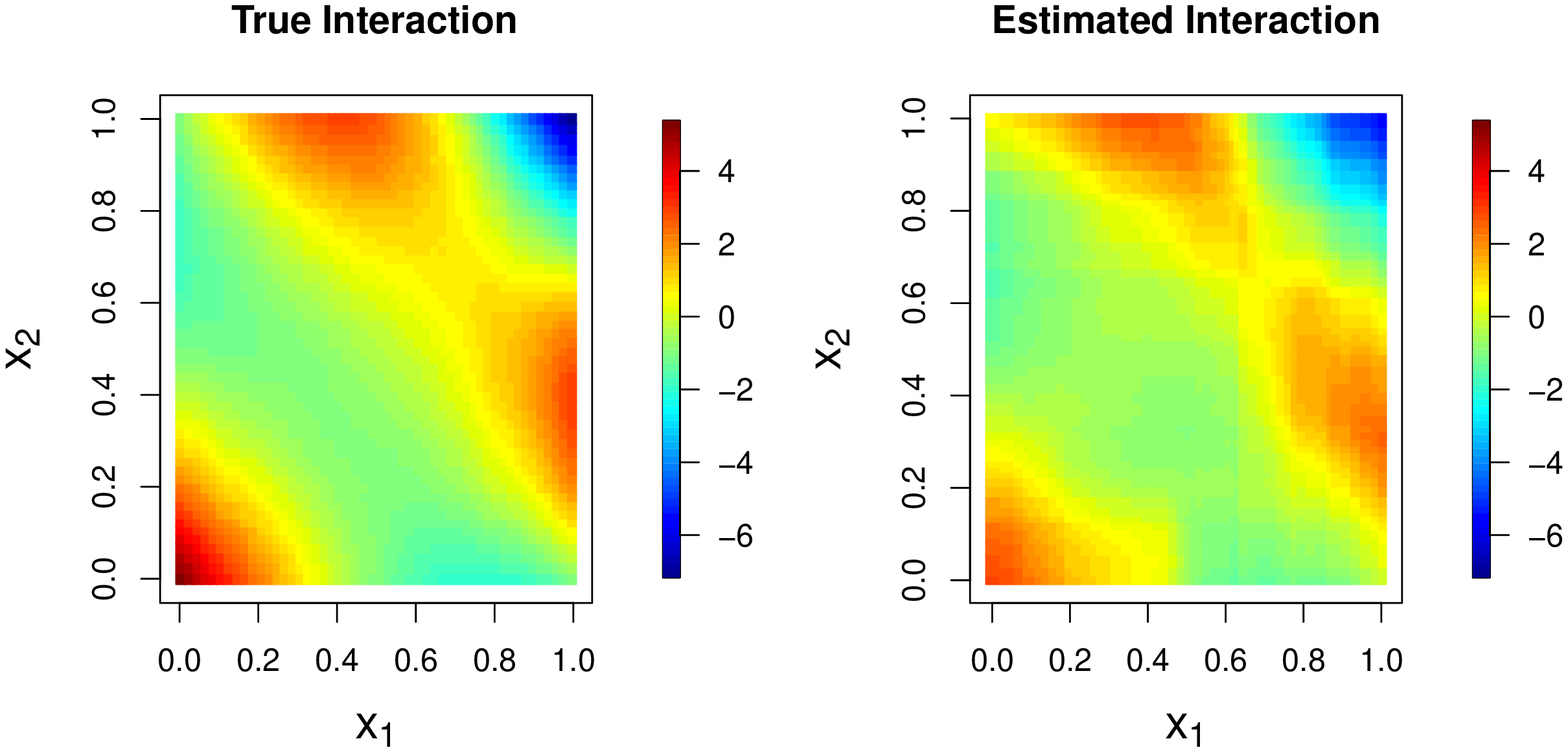}
    \caption{Two-way ALE plots for the actual Friedman function (left) and for the BPPR model of the Friedman function (right).}
    \label{fig:friedman_ALE_12}
\end{figure}

Recall that the only two-way interaction in the Friedman function is between $x_1$ and $x_2$, denoted $x_1$:$x_2$. \Cref{fig:friedman_ALE_12} shows the true two-way marginal effect of $x_1$:$x_2$ plotted next to the BPPR estimate, which match well, especially considering the relatively small sample size. The BPPR model does estimate the interactions $x_1$:$x_5$ and $x_2$:$x_5$ to be nonzero, but these estimates are small in magnitude and the corresponding error bounds include zero across more than $99\%$ of the domain.

\begin{figure}[h]
    \centering
    \includegraphics[width = 4.25in, keepaspectratio]{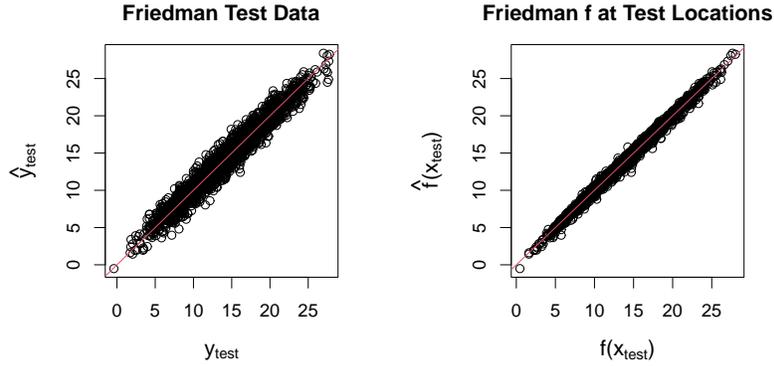}
    \caption{Predicted vs. actual values for test responses generated from the Friedman function (left) and the Friedman function evaluated at the test locations (right).}
    \label{fig:friedman_pred_v_actual_test}
\end{figure}

Finally, \Cref{fig:friedman_pred_v_actual_test} displays predicted values at 2,000 out-of-sample input locations, plotted against the out-of-sample response (left) and against the true Friedman function $f(\bm x)$ at these locations (right), with the one-to-one line in red. Predictions are highly accurate, with an out-of-sample RMSE of $1.05$ for the noisy response, and $0.38$ for $f(\bm x)$. Indeed, in our bake-off to follow, we find that BPPR provides the most accurate model of the Friedman function. Ninety-five percent prediction intervals for the noisy response cover 94.9\% of observed values, and 95\% credible intervals for $f(\bm x)$ cover 91.1\% of the true values, demonstrating good uncertainty quantification.

\subsection{Surge Data}\label{sec:surge}

We now demonstrate BPPR on a computer model emulation problem, where the simulated response is the maximum water level at a specific location during a storm surge resulting from a hurricane near the Delaware Bay (on the northeast seaboard of the United States) in the year 2100, as generated by SLOSH (see \cite{slosh} for details). Five inputs control characteristics of the hurricane at landfall (hurricane heading, velocity of the eye, maximum wind speed, minimum pressure, and location of landfall as a distance along the coast) and a sixth input is the sea level rise in the year 2100. Some of the simulated hurricanes result in no flooding at the location of interest; hence a large number of the responses are exactly zero.

Our primary goal is to closely emulate the computer model so that predictions at new inputs settings are precise, with accurate uncertainty quantification. To evaluate model performance, we randomly split the dataset into a training set with 3,000 observations and a test set with 1,000 observations. We fit the model with all hyperparameters set to their recommended defaults, and run the MCMC algorithm for 150,000 iterations to ensure convergence, which takes a little over 2 hours. The algorithm settles on $M = 31$ ridge functions for most posterior draws. Satisfied with convergence of $\sigma$ and $M$ (see the supplemental materials for details), we use the final 10,000 draws for posterior inference.

\begin{figure}[h]
    \centering
    \includegraphics[width = 4.25in, keepaspectratio]{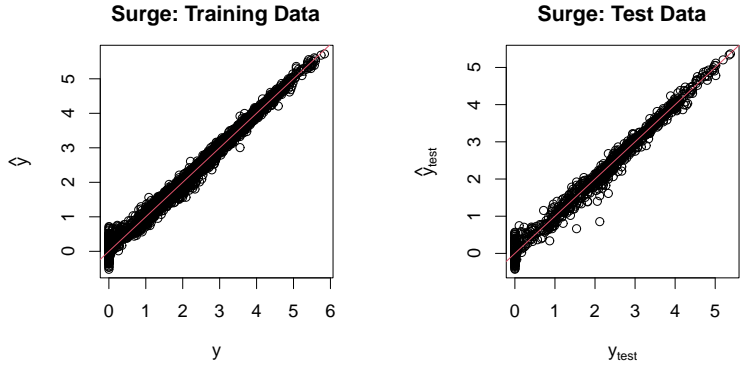}
    \caption{Predicted vs. actual values for the Surge data, with the training set on the left and the test set on the right.}
    \label{fig:surge_pred_v_actual}
\end{figure}

The left panel of \cref{fig:surge_pred_v_actual} evaluates the model fit to the training data, with observed response on the horizontal axis, fitted values on the vertical axis, and the one-to-one line plotted in red. In-sample RMSE is 0.139 and in-sample coverage of 95\% prediction intervals is 0.947. The right panel evaluates accuracy for the test set, which appears similar to that of the training data, with a few minor exceptions. RMSE for the test set is 0.165, and coverage is 0.924. As we will show in our bake-off, BPPR is the best emulator we've found for the Surge data in terms of out-of-sample predictive accuracy.

\begin{figure}[h]
    \centering
    \includegraphics[width = 4.05in, keepaspectratio]{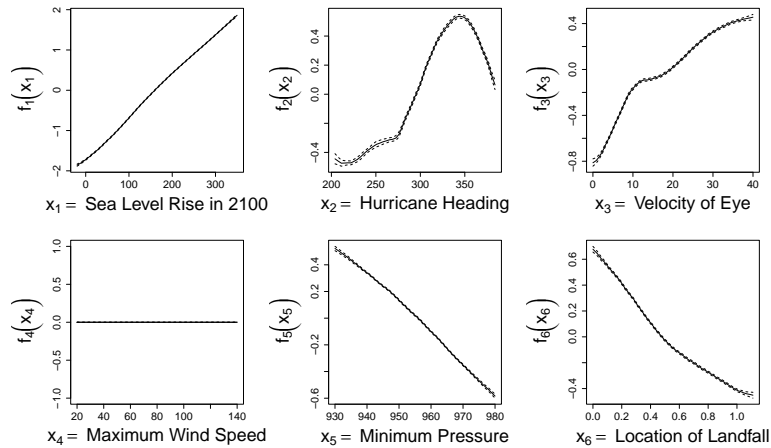}
    \caption{One-way ALE plots for the BPPR model of the surge data.}
    \label{fig:surge_ALE}
\end{figure}

\begin{figure}[h]
    \centering
    \includegraphics[width = 3.75in, keepaspectratio]{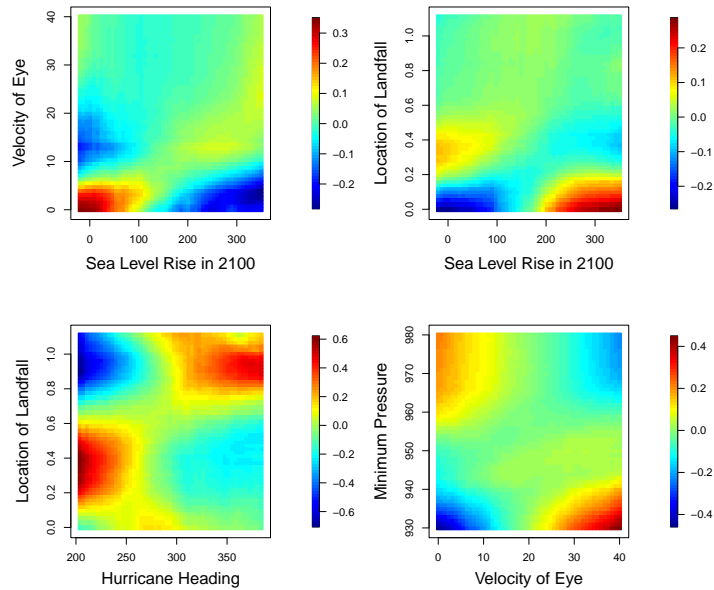}
    \caption{Two-way ALE plots for the BPPR model of the surge data.}
    \label{fig:surge_ALE2}
\end{figure}

It is also of interest to infer and verify known physical properties of the underlying process. To this end, \cref{fig:surge_ALE} displays ALE plots of all main effects, and \cref{fig:surge_ALE2} shows ALE plots of selected two-way interactions. From these, we see that maximum wind speed appears to play no role in the response function, since its estimated marginal effect is zero. Upon further investigation, we discovered that SLOSH does not use max wind speed because it is highly correlated with min pressure, implying that BPPR correctly discovered an inert variable here. All other variables have more substantial effects, with sea level rise having the largest marginal effect. We estimate that water level increases in an approximately linear fashion with sea level rise and that it decreases linearly with pressure, whereas velocity has a nonlinear relationship with water level, and we estimate a complicated nonlinear interaction between hurricane location and heading.

\subsection{Multivariate Response}

Here we consider two test functions used to simulate multivariate responses: (1) the environmental spill function under the same settings used in section 4.6 of \cite{bass} and (2) a version of the Friedman function where the first variable is treated as a functional input (similar to section 4.4 of \cite{bass}).  These are two functions for which BMARS performs well using the BASS R package.  The environmental spill function simulates the release of a pollutant into an environment, with four inputs characterizing the release and diffusion rate, and multivariate response over time and space.  We consider six spatial locations and 20 time points, though in practice we stack space and time so that the model only makes use of a multivariate response with $D = 120$ dimensions.  We use a training set of $n=$ 1,000 model evaluations, with no added noise.  For the Friedman function, we evaluate the first variable (treated as the functional input) on a grid of $D=50$ points, use a training set of $n=200$ function evaluations, and add standard Gaussian noise.  We also add five inert variables, for a total of nine inputs.  Hence, (1) demonstrates the method with $n=$ 1,000 noiseless function evaluations, with $p=4$-dimensional input and $D=120$-dimensional output, while (2) demonstrates the method with $n=200$ noisy function evaluations, with $p=9$-dimensional input (though only four matter) and $D=50$-dimensional response.

\begin{figure}[h]
    \centering
    \includegraphics[width = 3.05in, keepaspectratio]{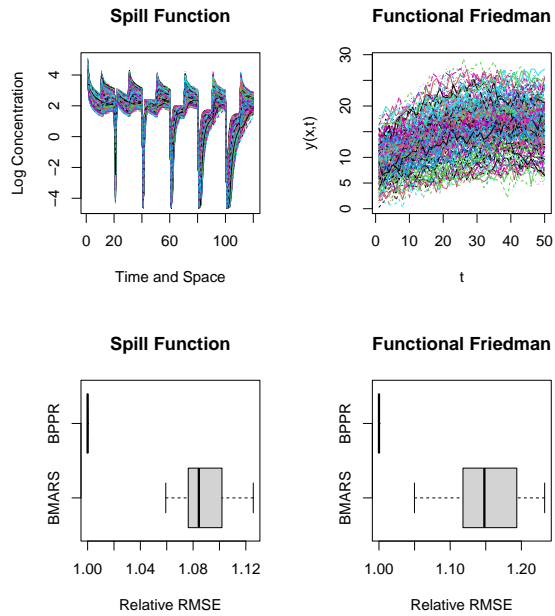}
    \caption{Multivariate response BMARS vs. BPPR for two different test functions.}
    \label{fig:compare_multi}
\end{figure}

\Cref{fig:compare_multi} shows the training responses for the two test functions.  We compare holdout prediction performance for our BPPR method used in PCA space to BMARS used in PCA space, where R code for the latter is detailed in \cite{bass}.  The boxplots in \cref{fig:compare_multi} are obtained by repeating the data generation and training 20 times for each test function.  In each training setting, the two methods (with default settings) are used to train a model using $D^-=15$ orthogonal basis functions and then predict at 1000 held out model evaluations.  The posterior predictive mean is used to obtain RMSE in each case, and the boxplots show relative RMSE (rescaled by the minimum for each training set).  Minimum RMSE was between (0.006, 0.008) for the spill function and (0.155, 0.190) for the Friedman function, indicating that both models fit the data well.  However, BPPR outperforms the BMARS model in all cases.  

\subsection{Bake-off}

In this section we compare the performance of BPPR against seven other state-of-the-art regression methods, including PPR, RF, LASSO, BART, BMARS, LAGP, and the Gramacy-Lian version of SIM, on an array of 20 simulated settings and 23 real datasets. We acknowledge that there are many useful function estimation methods, but this subset spans a wide range of likelihood-free and Bayesian methods, some of which are similar to BPPR while others are markedly different. For the likelihood-free methods, we perform a small cross-validation algorithm on each training set to optimize key hyperparameters, namely the number $M\in\{3, 5, 15, 20\}$ of ridge functions for PPR, the number $M\in\{50, 100, 200, 500\}$ of trees for RF, and the penalization parameter for LASSO, which we optimize using the cv.glmnet() function from the glmnet package in R. After cross-validation, we fit the full training set using the optimized values. Because the probabilistic methods require a longer run time, we use default hyperparameters for these, with no cross-validation on the training set. For BPPR, we collect a total of 10,000 posterior samples and use the last 1,000 of these to obtain predictions (note that we repeat the analyses in \cref{sec:friedman} and \cref{sec:surge}, but with these default settings), which makes BPPR's computation time comparable to that of BART and BMARS.

After each model has been fit to a training set of $n$ datapoints, predictions are made for a test set of size $n_{test}$. To evaluate performance, relative RMSE for the test set---normalized by the minimum RMSE across all models---is calculated for each model. This process is repeated for 20 different train/test splits.

\subsubsection*{Simulated Scenarios}

For the simulated datasets, training features $\bm x_1,\dots,\bm x_n$ are simulated, and corresponding Gaussian responses $y_1,\dots,y_n$ are generated with mean $f(\bm x_i)$ and standard deviation $\sigma$ for 15 different closed-form functions $f:\mathbb{R}^p\to\mathbb{R}$ with differing values of $p$ and $\sigma$, corresponding to various Signal-to-Noise Ratios (SNR). Each of these functions takes $p^{(\mathbb{R})}$ real-valued features as inputs, which are generated independently and uniformly from $[0, 1]^{p^{(\mathbb{R})}}$; in some cases this includes $r$ variables which are inert with respect to the response function. In addition, three of the functions also require the input of a categorical variable with $L$ levels, which is simulated independently and uniformly from $\{\mathcal{C}_1,\dots,\mathcal{C}_L\}$, and coded as $p^{(d)} = L - 1$ dummy variables $d_1,\dots,d_{p^{(d)}}$, bringing the total number of features to $p = p^{(\mathbb{R})} + r + p^{(d)}$. Test datasets are generated in the same way, with $n_{test} = n$. The names of the response functions, along with the values of $n$, $p$, $r$, $p^{(d)}$, and SNR used in the simulations, are given in figs. \ref{fig:sim_rmse1} and \ref{fig:sim_rmse2}, and sources for these functions are cited in the supplemental materials. For five of the functions, we try a second value of $(n, r, \text{SNR})$ so that we examine 20 simulation scenarios in all.

\begin{figure}
    \centering
    \includegraphics[width = 3in, keepaspectratio]{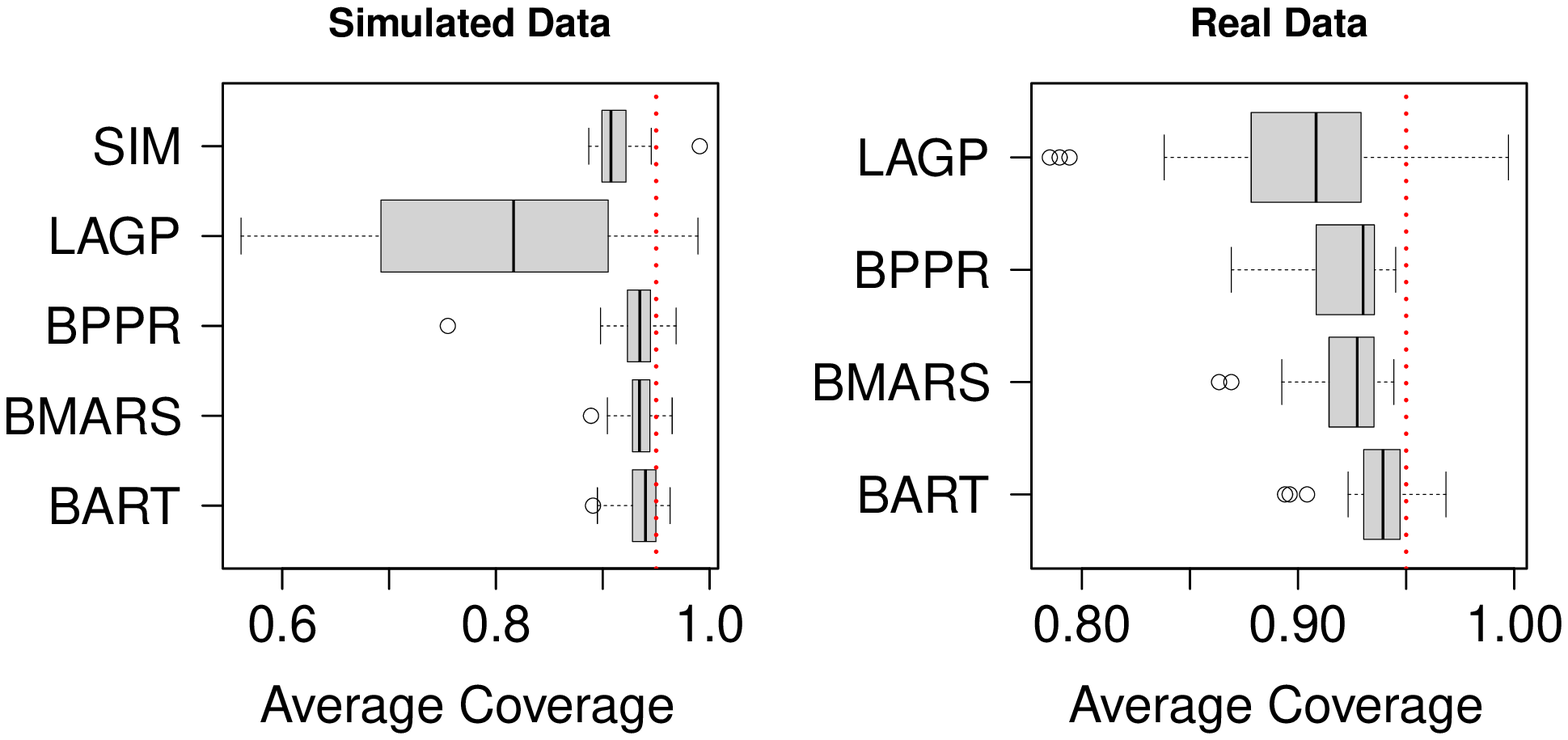}
    \caption{Average coverage of 95\% out-of-sample prediction intervals for each of the 20 simulation scenarios (left) and for each of the 23 real datasets (right).}
    \label{fig:cover}
\end{figure}

In terms of minimizing out-of-sample RMSE (figs. \ref{fig:sim_rmse1}-\ref{fig:sim_rmse2}), we find that BPPR performs best across simulation settings, with the lowest average RMSE for 13 of the 20 scenarios. BART comes in second by this measure, with the best accuracy in four cases; BMARS is most accurate in two cases; and SIM in one case. We evaluate uncertainty quantification in the left panel of \cref{fig:cover}, where empirical coverage is low in many cases for the LAGP method, and consistently slightly below nominal for the SIM model. BART credible intervals come closest to nominal coverage, on average, with BPPR and BMARS virtually tied for second except a single low outlier for BPPR.

\begin{figure}[h]
    \centering
    \includegraphics[width = 5.25in, keepaspectratio]{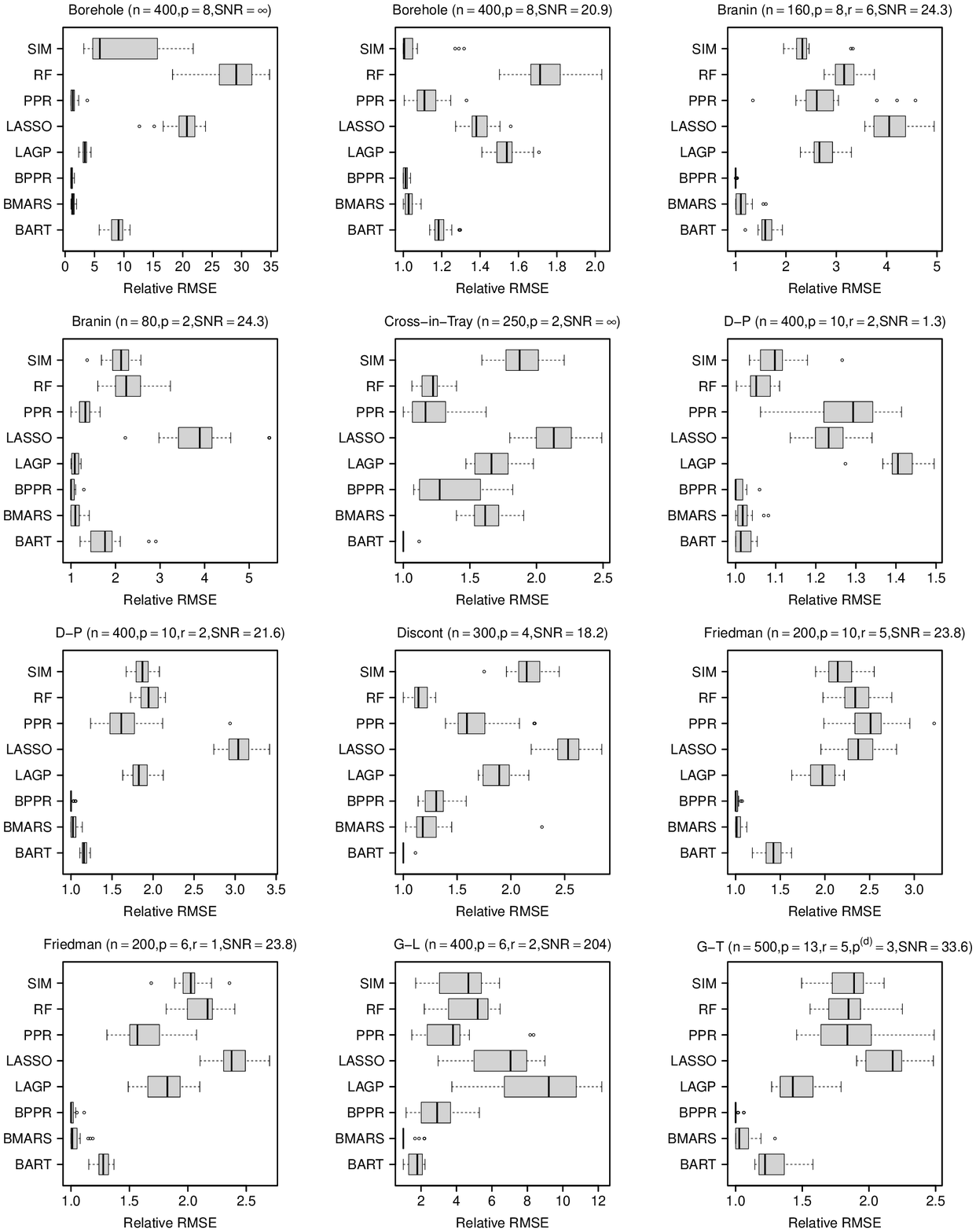}
    \caption{Relative RMSE for 12 of the simulation scenarios, with sample size $n$, number of features $p$, number of inert features $r$ (if any), number of dummy variables $p^{(d)}$ (if any), and SNR specified.}
    \label{fig:sim_rmse1}
\end{figure}

\begin{figure}
    \centering
    \includegraphics[width = 5.25in, keepaspectratio]{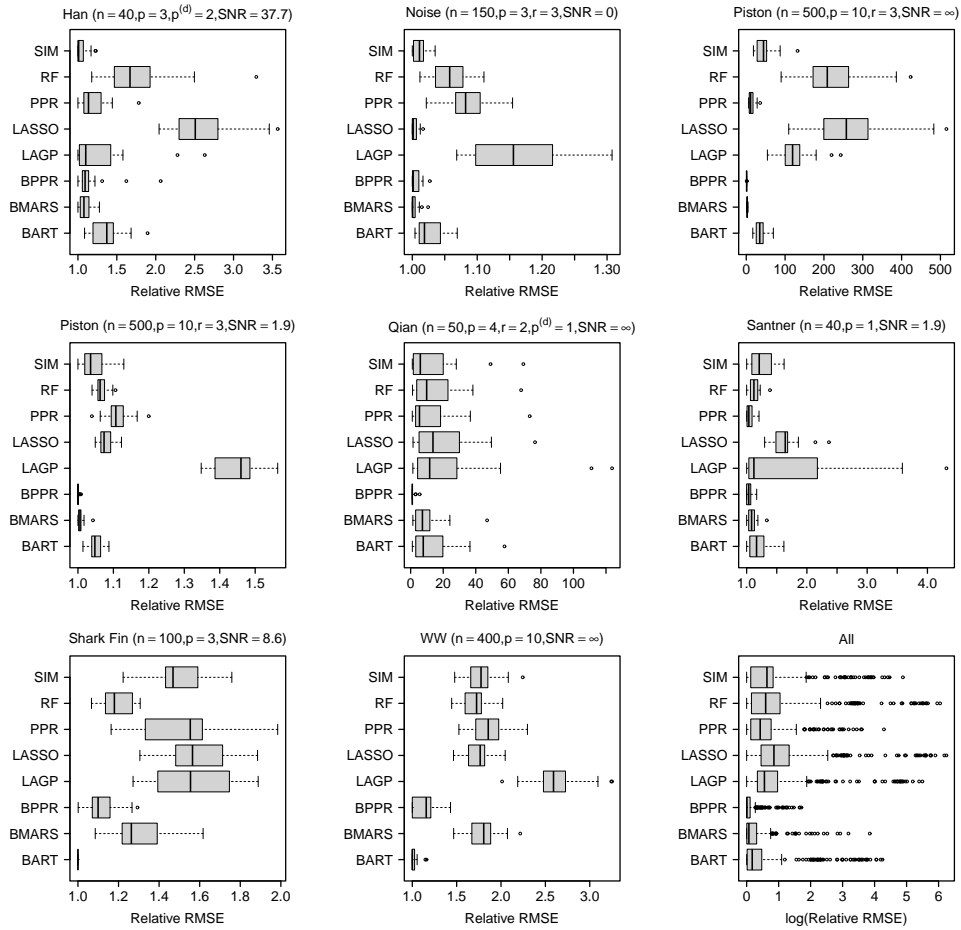}
    \caption{Relative RMSE for eight more simulation scenarios, and log Relative RMSE for all 20 simulation scenarios in the final panel, with summary statistics as in \cref{fig:sim_rmse1}.}
    \label{fig:sim_rmse2}
\end{figure}

\subsubsection*{Real Datasets}

For each real dataset, we randomly allocate 75\% of the data to a training set, with the remaining 25\% a test set. Seven of the datasets arise form real lab and computer experiments, and the other 16 are commonly used as benchmarks for statistical learning methods. These datasets are described and cited in the supplemental materials.

\begin{figure}
    \centering
    \includegraphics[width = 5.25in, keepaspectratio]{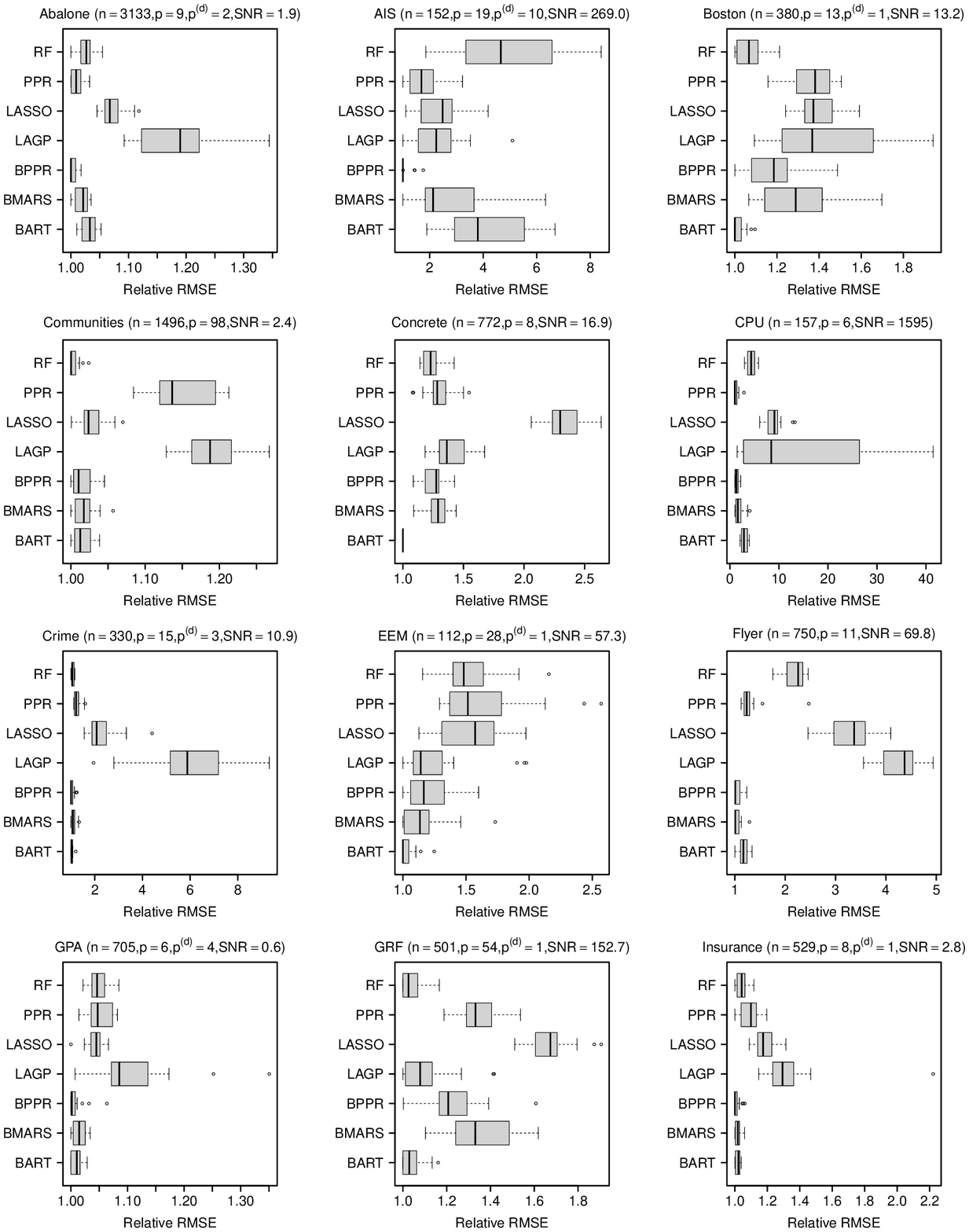}
    \caption{Relative RMSE for 12 of the real datasets. Sample size $n$ includes both training and test sets; $p$ and $p^{(d)}$ are as before; and SNR is estimated by $\frac{\hat{v}\{y_i\}_{i=1}^n - \min(MSE)}{\min(MSE)}$, where $\hat{v}\{y_i\}_{i=1}^n$ is the sample variance of $\{y_i\}_{i=1}^n$ and $\min(MSE)$ is the minimum out-of-sample MSE obtained by any method.}
    \label{fig:dat_rmse1}
\end{figure}

\begin{figure}
    \centering
    \includegraphics[width = 5.25in, keepaspectratio]{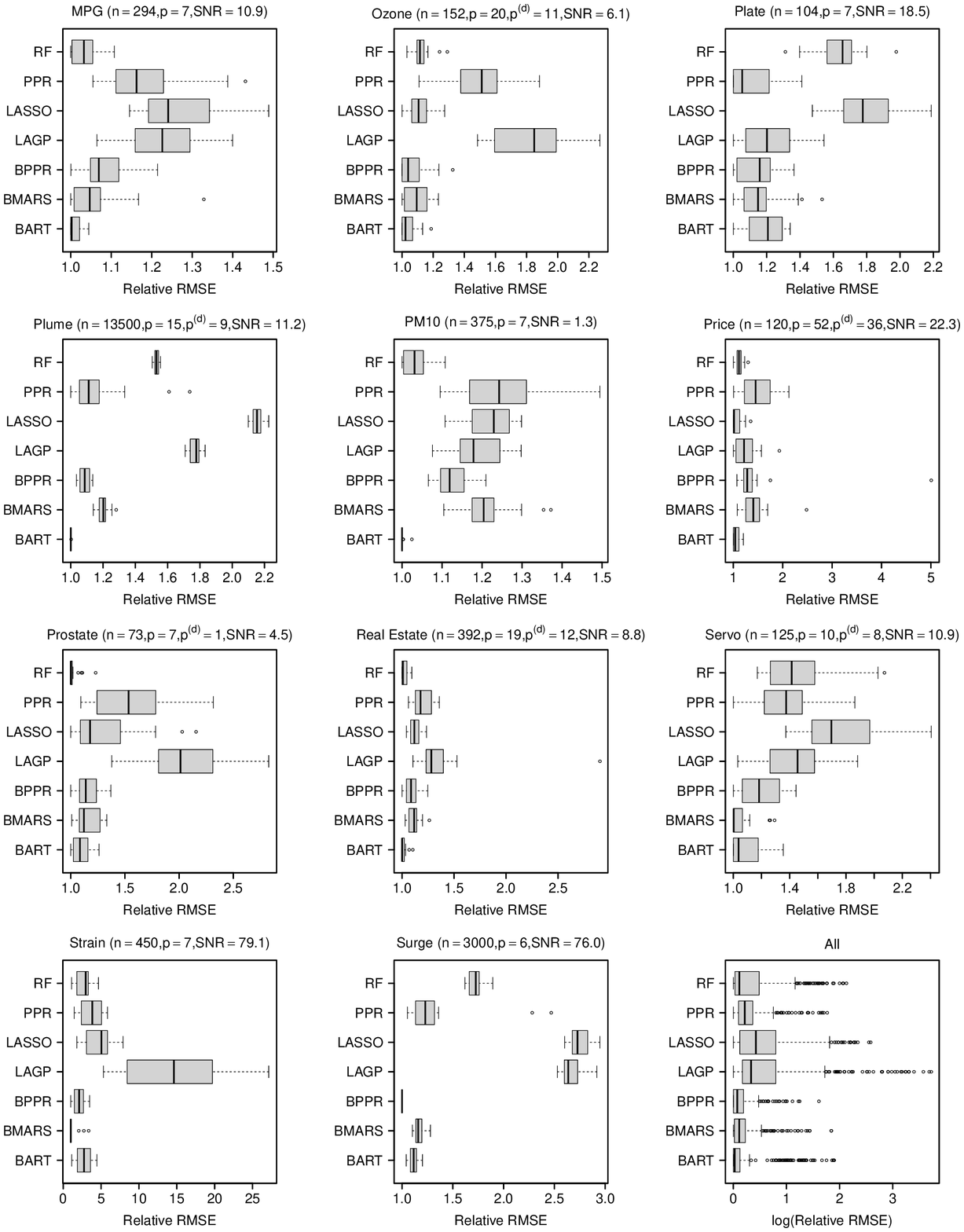}
    \caption{Relative RMSE for the 11 other real datasets, with summaries as in \cref{fig:dat_rmse1}.}
    \label{fig:dat_rmse2}
\end{figure}

Relative RMSE of test response predictions are shown in figs. \ref{fig:dat_rmse1} and \ref{fig:dat_rmse2}, with summaries of each dataset given in the panel titles, and coverage of 95\% prediction intervals is summarized in the right panel of \cref{fig:cover}. (Note the omission of SIM here because it is infeasible to fit this model to the larger datasets.) We find that BART performs best overall, with the lowest average RMSE for 10 of the 23 datasets, while BPPR is best for five datasets, and BMARS for three. Even when it is not the most accurate model, BPPR is still competitive, finishing second for eight datasets, third for six datasets, and never finishing worse than fifth. Overall, BART provides the most accurate uncertainty quantification, with BPPR and BMARS again virtually tied for second.

\section{Summary}\label{sec:summary}

To summarize, we have developed the first Bayesian version of PPR by combining various strategies from the SIM literature together with RJMCMC and some innovations that are novel to the PPR/SIM literature, including a modified natural spline expansion to form ridge functions, an efficient variable-selection process adapted from BMARS, and use of the power spherical distribution as a proposal for projection directions. We have also extended BPPR to accommodate categorical input variables and multivariate response. The result is a highly precise model with accurate uncertainty quantification for a wide array of real and simulated datasets, as shown in a bake-off where it often out-performs other state-of-the-art methods, and is competitive in all cases.

\clearpage
\singlespacing
\bibliographystyle{ba}
\bibliography{main.bib}

\onehalfspacing
\subsection*{Acknowledgments}
This material is based upon work supported by the National Science Foundation (CBET 1901845) and by Laboratory Directed Research and Development (LDRD) funding from Los Alamos National Laboratory. Any opinions, findings, and conclusions or recommendations expressed in this material are those of the authors and do not necessarily reflect the views of the National Science Foundation.

\end{document}


\maketitle
\onehalfspacing



\subsubsection*{Metropolis-Hastings Ratios for Death and Change Steps}
\label{sec:SMmh_dc}

In essence, the Metropolis-Hastings ratio for a death proposal is the reciprocal of the ratio for a birth proposal. For simplicity of expression, and using the notation given in the Death Proposal Algorithm (algorithm 3 in section 2.4 of the main paper), let
\begin{equation}
\begin{aligned}
    &\left(M, \bm B, \tau, a, \mathcal{J}, \bm\theta, t_0, \omega_a, \upsilon_j\right)\\
    &= \left(M^{(s-1)}, \bm B^{(s-1)}, \tau^{(s-1)}, a_{m^\star}^{(s-1)}, \mathcal{J}_{m^\star}^{(s-1)}, \bm\theta_{m^\star}^{(s-1)}, t_{m^\star 0}^{(s-1)}, \omega_a^{(s-1)}, \upsilon_j^{(s-1)}\right)
\end{aligned}
\end{equation}
represent the values from the previous iteration of the parameters involved in the death step. Then we have

\begin{equation}
\begin{aligned}
    \alpha^{(d)} = &\left[\frac{\pi(\bm y | \bm B^\star, \tau)}{\pi\left(\bm y | \bm B, \tau\right)}\right] \times \left[\frac{\pi(M^\star)}{\pi\left(M\right)\pi\left(a\right)\pi\left(\mathcal{J} | a\right)\pi\left(\bm\theta | \mathcal{J}, a\right)\pi\left(t_0 | \bm X \bm\theta\right)}\right]\\
    &\times \left[\frac{\omega_a^{(s)}\mathcal{W}_a\left(\mathcal{J} | \upsilon_1,\dots,\upsilon_p\right)\pi\left(\bm\theta | \mathcal{J}, a\right)\pi\left(t_0 | \bm X \bm\theta\right)/M}{1/M}\right]\\
    = &\left[\left(1 + \tau\right)^{\frac{K}{2}}\left(\frac{\bm y^\prime \bm y - \frac{\tau}{1 + \tau} \bm y^\prime \bm B^\star ({\bm B^\star}^\prime \bm B^\star)^{-1} {\bm B^\star}^\prime \bm y}{\bm y^\prime \bm y - \frac{\tau}{1 + \tau} \bm y^\prime \bm B (\bm B^\prime \bm B)^{-1} \bm B^\prime \bm y}\right)^{-n/2}\right] \times \left[\frac{\omega_a^{(s)} \times \mathcal{W}_a\left(\mathcal{J} \big| \upsilon_1,\dots,\upsilon_p\right)}{\lambda/(M A{p \choose a})}\right]
\end{aligned}
\end{equation}

Similarly, using notation given in the Change Proposal Algorithm (algorithm 4 in section 2.4 of the main paper) and letting $\left(\bm B, a, \mathcal{J}, \bm\theta, t_0\right) = \left(\bm B^{(s-1)}, a_{m^\star}^{(s-1)}, \mathcal{J}_{m^\star}^{(s-1)}, \bm\theta_{m^\star}^{(s-1)}, t_{m^\star 0}^{(s-1)}\right)$ represent the previous iteration's values of the parameters relevant to the change step, the Metropolis-Hastings proposal ratio for changing $\left(\bm\theta, t_0\right)$ to $\left(\bm\theta^\star, t_0^\star\right)$ in ridge function $m^\star$ is given by

\begin{equation}
\begin{aligned}
    \alpha^{(c)} = &\left[\frac{\pi(\bm y | \bm B^\star, \tau)}{\pi(\bm y | \bm B, \tau)}\right] \times \left[\frac{\pi(\bm\theta^\star | \mathcal{J}, a)\pi(t_0^\star | \bm X \bm\theta^\star)}{\pi(\bm\theta | \mathcal{J}, a)\pi(t_0 | \bm X \bm\theta)}\right]\\
    &\times \left[\frac{\mathcal{P}(\bm\theta_\mathcal{J} | \bm\theta^\star_\mathcal{J}, \kappa)\pi(t_0 | \bm X \bm\theta)}{\mathcal{P}(\bm\theta^\star_\mathcal{J} | \bm\theta_\mathcal{J}, \kappa)\pi(t^\star_0 | \bm X \bm\theta^\star)}\right]\\
    = &\left(\frac{\bm y^\prime \bm y - \frac{\tau}{1 + \tau} \bm y^\prime \bm B^\star ({\bm B^\star}^\prime \bm B^\star)^{-1} {\bm B^\star}^\prime \bm y}{\bm y^\prime \bm y - \frac{\tau}{1 + \tau} \bm y^\prime \bm B (\bm B^\prime \bm B)^{-1} \bm B^\prime \bm y}\right)^{-n/2},
\end{aligned}
\end{equation}

\noindent which is just the ratio of marginal likelihoods because we're using a symmetric proposal distribution for $\bm\theta^\star_\mathcal{J}$ (see eq. (17) in the main paper), and because the prior $\pi(\bm\theta | \mathcal{J}, a)$ is uniform (eq. (10) in the main paper).

\subsubsection*{Convergence of Markov Chains}

\begin{figure}
    \centering
    \includegraphics[width=3in]{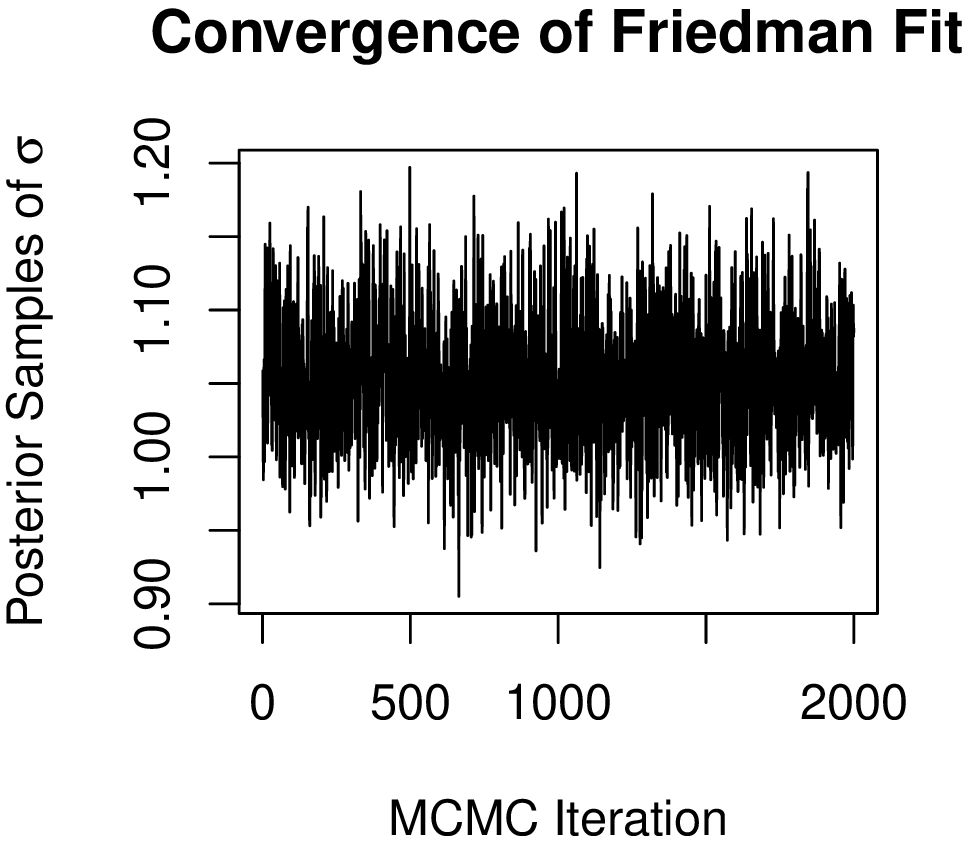}
    \caption{Traceplot of $\sigma$ in the BPPR model of the Friedman function, providing  evidence of convergence.}
    \label{fig:friedman_sd_resid_traceplot}
\end{figure}

Figure \ref{fig:friedman_sd_resid_traceplot} displays the traceplot for the 2,000 post-burn-in samples of $\sigma$ from the BPPR fit of the Friedman function in section 3.1 of the main text, which reveals no temporal trend and little evidence of sticking. The effective sample size of this chain is 1,815.9, and the split $\hat{R}$---calculated by splitting the samples into five sub-chains---is 1.001, which is close to the optimal value $1$. Furthermore, it is evident from the traceplot that the chain is centered close to the true value $\sigma = 1$. Altogether, these summaries provide strong evidence that the chain has converged.

\begin{figure}
    \centering
    \includegraphics[width=3in]{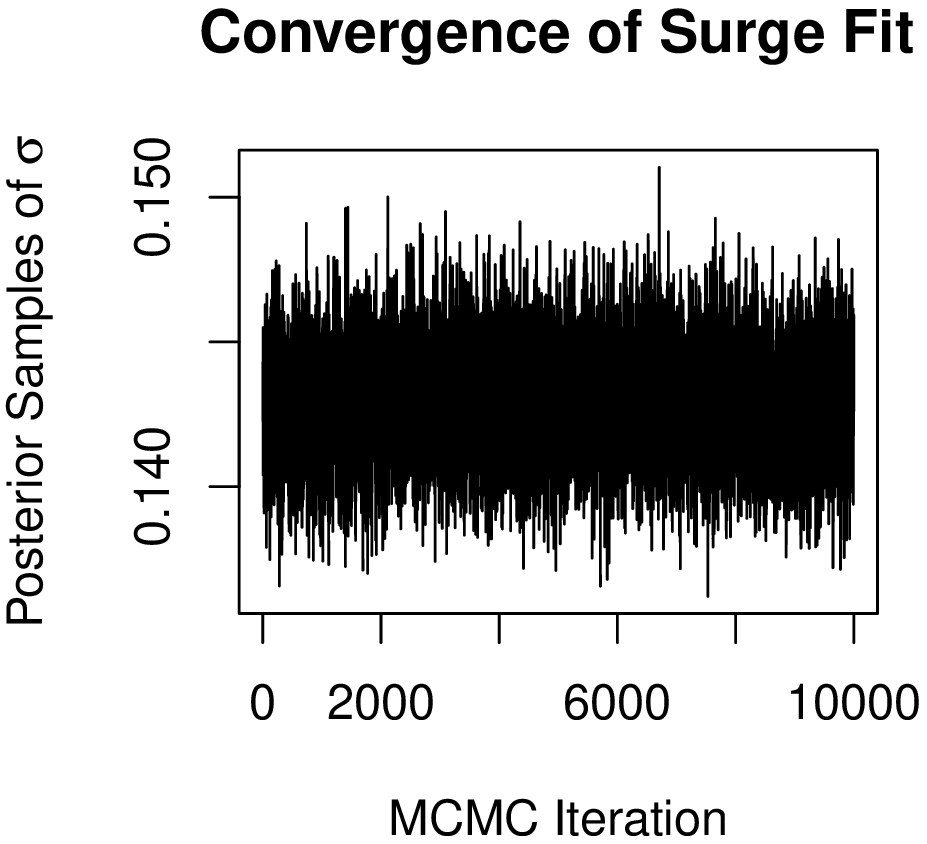}
    \caption{Traceplot of $\sigma$ in the BPPR model of the Surge data, providing  evidence of convergence.}
    \label{fig:surge_sd_resid_traceplot}
\end{figure}

Figure \ref{fig:surge_sd_resid_traceplot} displays the traceplot for the 10,000 post-burn-in samples of $\sigma$ from the BPPR fit of the Surge data in section 3.2 of the main text, which reveals little dependence between posterior draws. The effective sample size of this chain is 8,353.6, and the split $\hat{R}$---again calculated by splitting into five sub-chains---is 1.0029, which is again close to $1$. Once again, summaries provide substantial evidence of convergence.

\subsubsection{Description of Simulation Functions and Real Datasets}

First, we briefly cite our sources for the simulation functions from the bake-off in section 3.4 of the main text. The names and summary statistics for all functions are given in the panel titles of figs. 10 and 11 in the main text. The ``G-T'' function is from \cite{gt}, ``Shark Fin'' is a piecewise constant function we designed with tree-based models in mind, ``Noise'' indicates i.i.d. standard normal response with inert features only, and the remaining 12 functions are found in the Simon Fraser University library of test functions found in \cite{sfu}, where we closely follow the simulation scenarios described by the manuscripts cited therein except that we increase $n$ in a few instances.

Next, we describe and cite sources for the real datasets from the bake-off, with names and summaries of all datasets given figs. 13 and 14 in the main text. The first seven arise from real computer and lab experiments: the ``Surge'' dataset is described in section 3.2 of the main text. The ``EEM'' dataset measures Energy Expenditure per Minute, aggregated over time for individuals with various demographic characteristics walking/running at various speeds; additional inputs include data from a shoe-based analysis system worn by the subjects during data collection \citep{eem}. The inputs are similar, but more detailed, for the ``GRF'' dataset, where the response is maximum vertical Ground Reaction Force per stance, aggregated over many stances for a variety of individuals running at various speeds \citep{grf}. In the ``Flyer'' experiment, a computer model simulates launching a flyer into a plate, and the response is the velocity of the free surface of the plate due to the resulting shock wave under various simulation scenarios \citep{flyer}. Similarly, the ``Plate'' dataset simulates the deviation of the center of a tantalum plate after experiencing a shock wave at different input settings \citep{plate}. The simulated response for the ``Plume'' dataset is the log concentration of sulfur hexafluoride at a particular point in time and space after releasing the chemical upwind, with continuous inputs detailing the location, timing, and amount released; and categorical inputs indicating different meteorological settings \citep{plume}. Finally, the ``Strain'' dataset simulates the stress a material exhibits for particular input settings used to characterize the strength of a material \citep{strain}. The last 16 datasets are from commonly-used benchmark repositories. Specifically, the ``AIS'' dataset comes from the \textit{locfit} package, the ``PM10'' dataset comes from the \textit{truncSP} package, and the ``Boston Housing'' and ``Ozone'' datasets come from the \textit{mlbench} package in R; we obtained the ``Concrete,'' ``GPA,'' ``Insurance,'' ``Prostate,'' and ``Real Estate'' datasets from \cite{kutner}; and the remaining seven datasets come from the UCI machine learning repository \citep{uci}.

\singlespacing
\bibliographystyle{ba}
\bibliography{supplemental_materials.bib}

\onehalfspacing
\subsection*{Acknowledgments}
This material is based upon work supported by the National Science Foundation (CBET 1901845) and by Laboratory Directed Research and Development (LDRD) funding from Los Alamos National Laboratory. Any opinions, findings, and conclusions or recommendations expressed in this material are those of the authors and do not necessarily reflect the views of the National Science Foundation.